\documentclass[prb,showpacs,twocolumn,byrevtex]{revtex4}
\usepackage{amsmath}
\usepackage{graphicx}
\usepackage{float}
\usepackage{bm}
\usepackage{multirow}
\usepackage{dcolumn}

\newcommand{\icm}{\ensuremath{\textrm{cm}^{-1}}} % cm-1 

\newcolumntype{.}{D{.}{.}{-1}}
\begin{document}

%\linenumbers

\bibliographystyle{apsrev}

\title{Elastic and magnetic effects on the infrared phonon spectra of MnF$_2$}
\author{R. Schleck}
\affiliation{Laboratoire de Physique et \'Etude des Mat\'eriaux, LPEM, ESPCI, CNRS UPR5 Universit\'e Pierre et Marie Curie, 10 rue Vauquelin, F-75231 Paris Cedex 5, France}

\author{Y. Nahas}
\affiliation{Laboratoire de Physique et \'Etude des Mat\'eriaux, LPEM, ESPCI, CNRS UPR5 Universit\'e Pierre et Marie Curie, 10 rue Vauquelin, F-75231 Paris Cedex 5, France}

\author{R. P. S. M. Lobo}
\affiliation{Laboratoire de Physique et \'Etude des Mat\'eriaux, LPEM, ESPCI, CNRS UPR5 Universit\'e Pierre et Marie Curie, 10 rue Vauquelin, F-75231 Paris Cedex 5, France}

\author{J. Varignon}
\affiliation{CRISMAT, ENSICAEN-CNRS UMR~6508, 6~bd. Mar\'echal Juin, 14050 Caen, France}

\author{M. B. Lepetit}
\affiliation{CRISMAT, ENSICAEN-CNRS UMR~6508, 6~bd. Mar\'echal Juin, 14050 Caen, France}

\author{C. S. Nelson}
\affiliation{National Synchrotron Light Source, Brookhaven National Laboratory, Upton, NY 11973, USA}

\author{R. L. Moreira}
\affiliation{Departamento de F\'isica, ICEx, Universidade Federal de Minas Gerais, CP 702, Belo Horizonte MG 31120-970, Brazil}

%\author{J-Y Gesland}
%\affiliation{Universit\'e du Maine}

\date{\today}
\begin{abstract}
We measured the temperature dependent infrared reflectivity spectra of MnF$_2$ between 4~K and 600~K. We show that the phonon spectrum undergoes a clear renormalization at $T_N$. The \textit{ab-initio} calculation we performed on this compound accurately predicts the magnitude and the direction of the changes in the phonon parameters across the antiferromagnetic transition, showing that they are mainly induced by the magnetic order. In this material, we found that the dielectric constant is mostly from phonon origin. The large change in the lattice parameters with temperature seen by X-ray diffraction as well as the A$_{2u}$ phonon softening below $T_N$ indicate that magnetic order induced distortions in MnF$_2$ are compatible with the ferroelectric instabilities observed in TiO$_2$, FeF$_2$ and other rutile-type fluorides. This study also shows the anomalous temperature evolution of the lower energy $E_u$ mode in the paramagnetic phase, which can be compared to that of the $B_{1g}$ phonon seen by Raman spectroscopy in many isostructural materials. This was interpreted as being a precursor of a phase transition from rutile to CaCl$_2$ structure which was observed under pressure in ZnF$_2$.
\end{abstract}
\maketitle

%
% Introduction
%
\section{Introduction}

In magnetoelectric multiferroic materials, ferroelectricity coexists with a magnetic order. Based on the origin and strength of the coupling between the ferroelectric and the magnetic order parameters, these materials can be divided up into two general classes.\cite{KHOMSKII2009} In type-I multiferroics, such as BiFeO$_3$,\cite{Smolenskii1959,Kiselev1963} ferroelectric and (anti)ferromagnetic transitions are independent and weakly coupled. In type-II multiferroics, such as TbMnO$_3$,\cite{Kimura2003} ferroelectricity is a consequence of the magnetic ordering and a strong magneto-electric coupling is present. In the latter compounds, the interaction between magnetic ordering and the lattice may generate the structural distortions leading to the appearance of a permanent electrical polarization. Although the microscopic origins for this magnetoelectric coupling are still under debate, the appearance of an electrical dipole moment has important consequences on the polar, infrared active, phonon spectra. Very few studies of the phonon changes in these type-II multiferroics exist to date. Nevertheless, Schmidt \textit{et al.}\cite{Schmidt2009} showed that in TbMnO$_3$ phonon parameters are renormalized by about 1--2\% and that the changes are linked to $T_N$ rather than the ferroelectric transition.

Before tackling the structurally complex type-II multiferroics, it is interesting to see the effect of magnetic ordering on the infrared phonon response. Several infrared studies on spinels,\cite{Wakamura1988,Rudolf2007,Sushkov2005} showed that the magnetic ordering leads to infrared phonon splitting. However, the phonon splitting has systematically been attributed to the magnetic frustration present in these compounds, a vision supported by \textit{ab-initio} calculations.\cite{Fennie2006} Previous studies on antiferromagnetic MnO (Ref.~\onlinecite{RUDOLF2008}) and CoO (Ref.~\onlinecite{KANT2008}) showed a strong phonon renormalization at $T_N$. However, both materials also undergo a structural phase transition from cubic to rhombohedral (MnO) or tetragonal (CoO) at $T_N$. These phase transitions make it harder to separate what is the pure magnetic effect on phonons from what is due to the structural symmetry change.

In this perspective, manganese fluoride (MnF$_2$) emerges as a system of choice, once it is a very well characterized commensurate antiferromagnet with $T_N = 68$~K.\cite{KATSUMATA2000} Its simple paramagnetic rutile structure\cite{STOUT1954} ($P_{4_2}$/mnm or D$_{4h}^{14}$) remains the same below $T_N$, where the spins align antiferromagnetically along the $D_4$ axis. Hence effects on the phonon spectra have a magnetic origin.

Temperature dependent Raman spectra of MnF$_2$ show phonon frequency changes at $T_N$.\cite{LOCKWOOD1988} As this material has an inversion center, Raman active phonons are not infrared active and \textit{vice versa}. Therefore, detailed knowledge of infrared phonons in MnF$_2$ is paramount to grasp changes that would affect dielectric ordering below $T_N$. We expect that these data on MnF$_2$ help to set a baseline to understand the effect on phonons of the magnetic transitions in type-II multiferroic materials.

To date, only the room temperature infrared phonon spectra have been measured for MnF$_2$.\cite{WEAVER1974} According to dilatometric measurements\cite{GIBBONS1959} MnF$_2$ lattice parameters are renormalized across the antiferromagnetic transition. This magnetostrictive effect is the hallmark of the coupling between dipole excitations and magnetic order in this compound.

Although MnF$_2$ is a classical antiferromagnet, it is closely related to multiferroic materials. Its isostructural compound TiO$_2$ is a quantum paraelectric (incipient ferroelectric) as determined by infrared,\cite{Gervais1974} Raman and dielectric measurements,\cite{SAMARA1973} as well as \textit{ab-initio} calculations.\cite{MONTANARI2004} Similar responses, in particular renormalization of the phonon spectra, were observed in other rutile fluorides such as FeF$_2$,\cite{LOCKWOOD1983} ZnF$_2$,\cite{Giordano1988} and NiF$_2$.\cite{LOCKWOOD2002}

In this paper we show a detailed, temperature-dependent infrared study of the phonon spectra in MnF$_2$ along the $ab$-plane and the $c$-axis. Our results show that the phonon spectra have marked changes at $T_N$. The infrared data are complemented by low temperature X-ray diffraction and \textit{ab-initio} calculations. First principles calculations predict the proper phonon frequencies for both directions and find the correct frequency shifts at $T_N$. Our results show that the dielectric constant of MnF$_2$ is mostly from phonon origin. The large change in the lattice parameters with temperature and phonon softening in the antiferromagnetic phase suggest that MnF$_2$ distortions are compatible with the ferroelectric instabilities observed in TiO$_2$,\cite{MONTANARI2004} FeF$_2$,\cite{LOCKWOOD1983} and ZnF$_2$.\cite{Giordano1988}

%%%%%%%%%%%%%%%%%%%%%%%%%%%%%%%%%%%%%%%%%%%%%%%%%%%%%%%%%%%%%%%%%%%%%%%%%%%%%%%
%
% Methods
%
\section{Methods}
\label{Methods}

\begin{figure}[htb]
  \includegraphics[width=8cm]{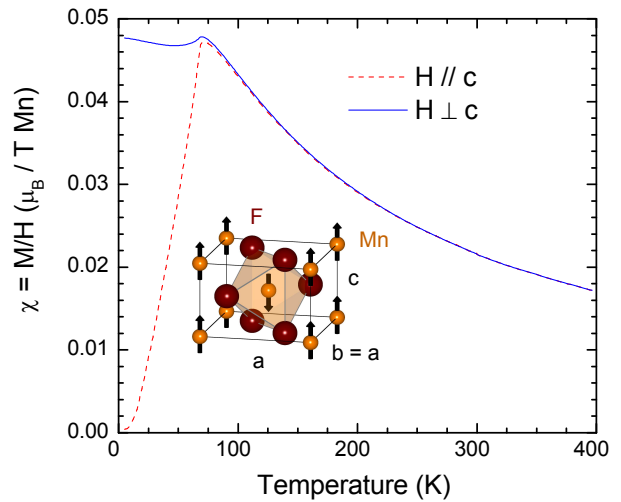} 
  \caption{(color online) Magnetic susceptibility of MnF$_2$ measured with $H = 1000$ gauss. The solid (blue) line is the data for $H \perp c$ and the dashed (red) line is obtained for $H \parallel c$. The diagram represents the lattice and magnetic structure of the rutile tetragonal phase, where $a = b = 4.874$~\AA\ and $c = 3.300$~\AA.}
  \label{magnetic}
\end{figure}

The MnF$_2$ single crystal used in this experiment was grown by the Czochralski method. X-ray and infrared measurements were done on different platelets cut from the same bulk and containing $ac$ and $ab$ planes. The $ac$ sample for optical measurements was polished with a 15$^\circ$ wedge to avoid interference fringes from the back surface reflectance. The measured faces for both optical samples were polished with 1$\mu$m diamond powder to a mirror like surface. Typical sample surface sizes are $5 \times 5$~mm$^2$. Their antiferromagnetic transition was measured on a Quantum Design MPMS-5 squid magnetometer with a magnetic field of 1000 gauss. Figure~\ref{magnetic} shows the magnetic susceptibility ($\chi$) measured with the field applied parallel and perpendicular to the c axis. The $H \parallel c$ curve shows a classical antiferromagnetic ordering behavior where $\chi$ decreases below $T_N$ reaching a vanishingly small value at $T = 0$~K. This indicates a negligible amount of impurities in this sample. When $H \perp c$, $\chi$ is dominated by a spin canting response below $T_N$.

We measured the infrared reflectivity spectra near normal incidence at 31 different temperatures between 4~K and 300~K, on the $ac$ plane sample with the electric field of light parallel and perpendicular to the $c$ direction ($D_4$ axis). To determine the absolute reflectivity, we used an \textit{in situ} gold overfilling technique.\cite{HOMES1993b} The sample is attached to a cold finger of an ARS Helitran cryostat and we measure its reflectivity with respect to a reference stainless mirror for all temperatures. We then evaporate a thin layer of gold on the sample producing a mirror with the same surface quality and shape as the sample. The reflectivity of this gold mirror is measured, at several temperatures, against the same stainless steel reference and the infrared spectrum is obtained by dividing the uncoated sample by its coated version. Finally, we multiply the result by the absolute reflectivity of gold to get the absolute reflectivity of the sample. The accuracy of the absolute reflectivity is better than 1\% and the relative error between different temperatures is of the order of 0.1\%. The far infrared (10--700~\icm) data were collected with a Bruker IFS113v interferometer. Higher frequency spectra (500--7500~\icm) were obtained with a Bruker IFS66v spectrometer. In the overlapping region the spectra agree within 0.5\%.

We also measured the reflectivity of the $ab$ sample, in the (100--700~\icm) range, at high temperatures (300--600~K) with a TS-1500 Linkam hot stage. No polarization was used in these measurements as $a$ and $b$ are equivalent axes. An Al mirror served as a reference. As no \textit{in situ} gold evaporation was used above room temperature, the data were corrected so that 300 K measurements in the cryostat and in the hot stage match.

X-ray diffraction measurements were carried out on the National Synchrotron Light Source beamline X21. A Si(111) double-crystal monochromator was used to set the incident energy of the beam to 11.5 or 13.5 keV, and a Pt-coated mirror focused the beam down to $\sim 1$ (vertical) by $2$ (horizontal) mm$^2$. The MnF$_2$ crystals were inserted in an Oxford superconducting magnet, which is mounted on a 2-circle diffractometer with a horizontal scattering geometry. A LiF(200) analyzer was used, and two parallel reflections were measured for both $ac$ and $ab$ plane samples to obtain the $a(=b)$ and c lattice parameters.

These measurements were complemented with \textit{ab-initio} density functional (DFT) calculations. In order to study the effects of the magnetic order on the phonon parameters, we performed the geometry optimization and phonon calculations for two different spin configurations. The first one is a true ferromagnetic (FM) configuration with all spins aligned. The second configuration is a pseudo-antiferromagnetic (AFM) configuration where nearest neighbor spins are anti-aligned, as shown in Fig.~\ref{magnetic} (up-down configuration). This configuration is only pseudo-antiferromagnetic since it is not an eigenfunction of the total spin operator. The true antiferromagnetic state is a superposition of the up-down and down-up configurations, as well as quantum fluctuations on them. Such a correct calculation is unfortunately not feasible for an infinite system. The calculations were done using the CRYSTAL06 package.\cite{CRYSTAL} Three different functionals were used for the calculations, namely LDA, and two hybrid functionals B3LYP\cite{B3LYP} and B1PW.\cite{B1PW} An atomic basis set of valence 2-$\zeta$ quality\cite{BaseMn} and small core pseudopotentials\cite{PseudoMn} were used for the Mn$^{2+}$ ions and an all electrons basis of 3-$\zeta$ quality was used for the F$^-$ ions.\cite{BaseF} As usual, the LDA functional underestimates the lattice parameters by a few percent while the two other functionals slightly overestimate them. The best fit is reached for the B1PW functional with an error of 0.2\% in the $a,b$ direction and 1\% in the $c$ direction. All \textit{ab-initio} results further presented in the paper will thus refer to the B1PW functional.

%%%%%%%%%%%%%%%%%%%%%%%%%%%%%%%%%%%%%%%%%%%%%%%%%%%%%%%%%%%%%%%%%%%%%%%%%%%%%%%
%
% Results
%
\section{Results}

\begin{figure}
  \includegraphics[width=8cm]{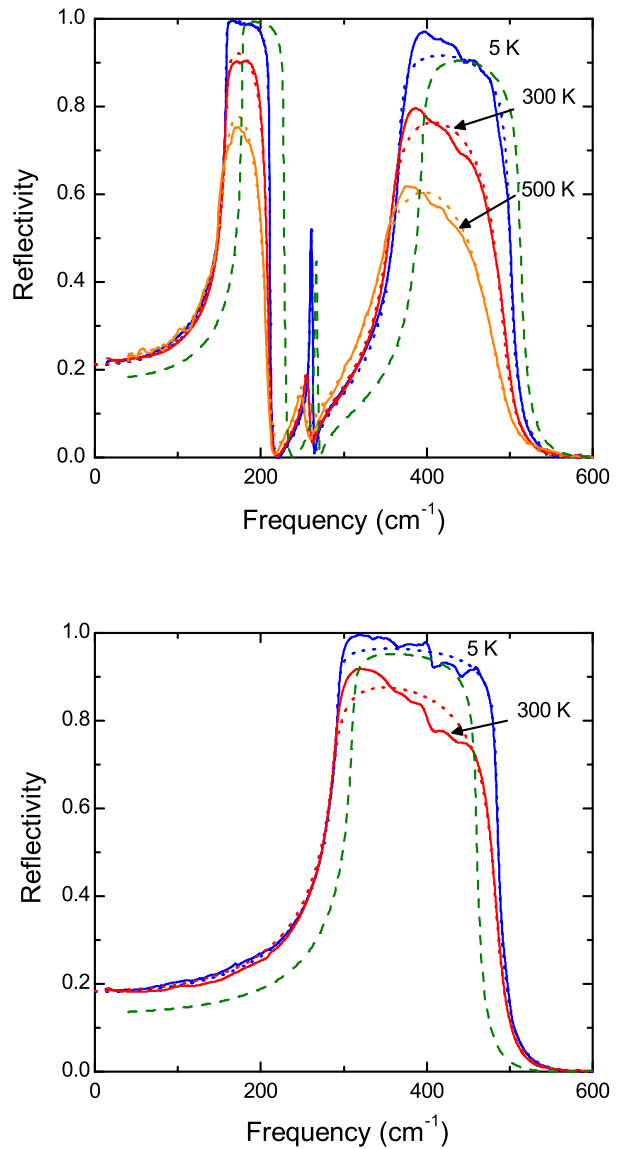}
  \caption{(color online) Reflectivity with light polarized in the $ab$-plane (top panel) and along the $c$-axis (bottom panel) for MnF$_2$. The solid lines are experimental results and the dotted lines are Lorentz fits using Eq.~\ref{lorentz}. In each panel, the dashed line is the reflectivity predicted by \textit{ab-initio} calculations. These curves were generated by plugging into Eq.~\ref{lorentz} the first principles values for $\Omega_{TO}$ and $\Delta\varepsilon$ and using the $\gamma$ values obtained from the experimental fits at 5K. The \textit{ab-initio} values come from the AFM calculation.}
  \label{ReflAC}
\end{figure}

The solid lines in the top panel of Fig.~\ref{ReflAC} show the infrared reflectivity at selected temperatures with the electric field of light lying on the $ab$-plane. These spectra show three clear phonon peaks. The solid lines in the bottom panel are the reflectivity for $E \parallel c$ and show a single phonon along this direction. The infrared modes found agree with group theory predictions. Indeed, the irreducible representation decomposition for the rutile structure is $A_{1g} \oplus A_{2g} \oplus 2 A_{2u}\oplus B_{1g} \oplus 2B_{1u} \oplus B_{2g} \oplus E_g \oplus 4E_u$. Four of these modes are Raman active --- $A_{1g}$, $B_{1g}$, $B_{2g}$ and $E_g$ --- and four are IR active --- $ 3 E_u \oplus A_{2u}$. The $E_u$ modes are $xy$ degenerate and represent the $ab$ plane spectra. From low to high frequencies, we define these modes as $E_{u1}$, $E_{u2}$, and $E_{u3}$. The $A_{2u}$ mode has $z$ symmetry and is the only phonon along the $c$ axis.

To extract quantitative information from these data, we simulated the spectra with a multiple Lorentz oscillator model for the dielectric function:
%
%\begin{linenomath}
\begin{equation}
\varepsilon(\omega)=\varepsilon_\infty+\sum_k\frac{\Delta\varepsilon_k \Omega_{TO_k}^2}{\Omega_{TO_k}^2-\omega^2-i\gamma_k\omega},
\label{lorentz}
\end{equation}
%\end{linenomath}
%
where $\varepsilon_\infty$ is the contribution from electronic transitions to the dielectric function, and each phonon is described by a resonance frequency $\Omega_{TO_k}$, an oscillator strength $\Delta\varepsilon_k$, and damping $\gamma_k$. The reflectivity at normal incidence is given by $R = |1 - \sqrt{\varepsilon}|^2 / |1 + \sqrt{\varepsilon}|^2$.

Typical fitting results are shown as dotted lines in both panels of Fig.~\ref{ReflAC}. Overall, the model reproduces the data very well. However, modes $E_{u3}$ and $A_{2u}$, as well as $E_{u1}$ at high temperatures, have structures that cannot be reproduced by the Lorentz model. This structure is likely related to a breakdown of the harmonic approximation\cite{Sun2008} and/or two-phonon absorption,\cite{Benoit1988} even though we cannot rule out the presence of a small symmetry breaking lattice distortion. Nevertheless, in general, the parameters obtained through a Lorentz modeling of the data assuming a $D_{4h}^{14}$ symmetry are a very good first approximation.

To ascertain that this is the case with our data, we also used Kramers-Kronig transformations, which are model independent. These transformations require knowledge of the reflectivity in the full spectral range, from zero to infinity. As our measurement range is limited, we extrapolated the low frequency range as a constant reflectivity. For high frequencies we used a constant up to 80~000~\icm\ followed by a free-electron approximation ($R\propto\omega^{-4}$). The 80~000~\icm\ limit was chosen to avoid unphysical negative values (albeit within error bars) in the imaginary part of the dielectric function ($\varepsilon_2$).

\begin{figure}
  \includegraphics[width=8cm]{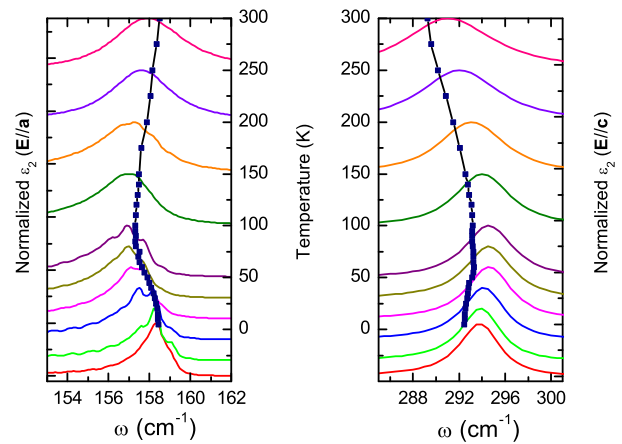}
  \caption{(color online) The left panel shows $\varepsilon_2$ spectra obtained by Kramers-Kronig transformation (solid curves) for phonon $E_{u1}$ and the right panel for phonon $A_{2u}$. For clarity, each $\varepsilon_2$ curve was normalized by its maximum value then shifted vertically by a value proportional to its temperature (the curves are placed so that their maxima coincide with the temperature scale shown). The squares are $\Omega_{TO}$ frequencies obtained from the Lorentz fits and are shown in a temperature \textit{vs.} frequency plot. In a perfect agreement between Lorentz and Kramers-Kronig analysis, each square would sit exactly on the maximum of a curve.}
  \label{KKFit}
\end{figure}

In Fig.~\ref{KKFit} we compare Kramers-Kronig results to Lorentz fit parameters. The left panel shows $\varepsilon_2$ for the $E_{u1}$ mode. For clarity, each curve was normalized by its maximum value and shifted vertically by an amount proportional to its temperature. The peak in $\varepsilon_2$ happens at the phonon resonance frequency. The squares are the $\Omega_{TO}$ frequencies obtained from the Lorentz fits to the data. Although a small difference in frequency is seen, both methods give the same temperature dependence and magnitude of the changes in the phonon features. The right panel presents the same comparison for the lone $A_{2u}$ phonon. This figure shows that, even though Eq.~\ref{lorentz} neglects anharmonic effects, its outcoming parameters are representative of the physical properties of MnF$_2$. We also checked that comparison of the Lorentz fits to other methods such as four-parameter simulations\cite{Gervais1974b} and multi-oscillator fits\cite{KUZMENKO2005} give the same results. Henceforth we will discuss our results using the Lorentz oscillator model alone.

We performed \textit{ab-initio} calculations to predict the values of infrared active phonon frequencies and oscillator strengths. The results of these calculations for the infrared active modes are shown as dashed lines in both panels of Fig.~\ref{ReflAC}. For reference, the results for all modes are given in the appendix. The \textit{ab-initio} frequencies are between 2 and 12\% higher than the measured ones. Except for the very weak $E_{u2}$ phonon, the predicted strenghs are underestimated by 4\% to 35\%. Nevertheless, the calculated spectra describe very well the overall phonon response. Table~\ref{table_params} summarizes the fitting parameters produced by Eq.~\ref{lorentz} at 5~K, 100~K and 300~K. It also shows the \textit{ab-initio} results for $\Omega_{TO}$ and $\Delta\varepsilon$ in the FM and AFM configurations defined in Sec.~\ref{Methods}.
\begin{table*}
\begin{center}
\caption{Lorentz fit parameters at 5~K, 100~K and 300~K and \textit{ab-initio} results for the phonon frequencies and oscillator strengths. Units for $\Omega_{TO}$ and $\gamma$ are \icm. FM and AFM configurations are defined in Sec.~\ref{Methods}. Fitted values for $\varepsilon_\infty$ are 2.16 for the $ab$-plane and 2.25 along the $c$ axis.\label{table_params}}
\begin{ruledtabular}
\begin{tabular}{c..................}
\multicolumn{3}{c}{} & \multicolumn{1}{c}{5K} & \multicolumn{3}{c}{} & \multicolumn{1}{c}{100K} & \multicolumn{3}{c}{} & \multicolumn{1}{c}{300K}& \multicolumn{2}{c}{} & \multicolumn{2}{c}{AFM}	& \multicolumn{1}{c}{} & \multicolumn{2}{c}{FM}\\
\cline{3-5}\cline{7-9}\cline{11-13}\cline{15-16}\cline{18-19}
\multicolumn{1}{c}{} & \multicolumn{1}{c}{} & \multicolumn{1}{c}{$\Omega_{TO}$ }&	\multicolumn{1}{c}{$\Delta\epsilon$} &\multicolumn{1}{c}{ $\gamma$ }& \multicolumn{1}{c}{} & \multicolumn{1}{c}{$\Omega_{TO}$ }&	\multicolumn{1}{c}{$\Delta\epsilon$} &\multicolumn{1}{c}{ $\gamma$ }& \multicolumn{1}{c}{} & \multicolumn{1}{c}{$\Omega_{TO}$} 	& \multicolumn{1}{c}{$\Delta\epsilon$}	& \multicolumn{1}{c}{$\gamma$	}&  \multicolumn{1}{c}{} & \multicolumn{1}{c}{$\Omega_{TO}$}	& \multicolumn{1}{c}{$\Delta\epsilon$}&  \multicolumn{1}{c}{} & \multicolumn{1}{c}{$\Omega_{TO}$}& \multicolumn{1}{c}{$\Delta\epsilon$}\\
\hline
$A_{2u}$ & & 292.5 & 3.93 & 4.37 & & 293.2 & 3.92 & 6.42 & & 289.6 & 4.01 & 16.01 & & 311.0 & 2.53 & & 315.1 & 1.74 \\
\\ 
$E_{u1}$ & & 158.4 & 3.56 & 0.25 & & 157.3 & 3.66 & 0.96 & & 158.5 & 3.59 & 3.48 & & 178.2 & 2.68 & & 175.2 & 2.71 \\
$E_{u2}$ & & 259.8 & 0.08 & 1.56 & & 258.9 & 0.08 & 2.48 & & 255.1 & 0.09 & 6.88 & & 265.7 & 0.06 & & 265.5 & 0.06 \\
$E_{u3}$ & & 367.0 & 1.48 & 7.00 & & 363.3 & 1.50 & 9.47 & & 357.9 & 1.51 & 21.70 & & 394.6 & 1.20 & & 387.7 & 1.17 \\
\end{tabular}
\end{ruledtabular}
\end{center}
\end{table*}

\begin{figure}
  \includegraphics[width=8cm]{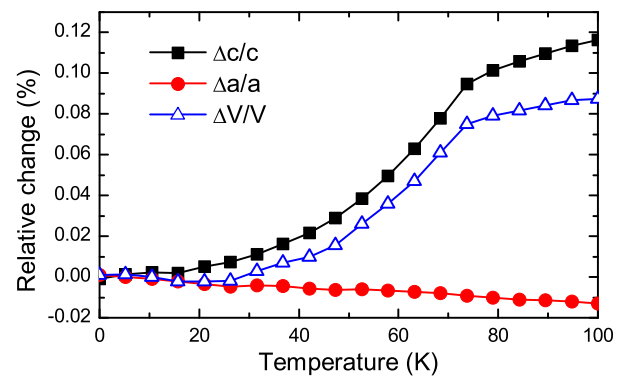}
  \caption{(color online) Relative changes in $a$ and $c$ lattice parameters measured by x-ray diffraction. We also show the resulting relative variation of the unit-cell volume (open triangles).}
  \label{xray}
\end{figure}

X-ray diffraction results are presented on Fig~\ref{xray} where we plot the relative change of both lattice parameters as a function of temperature. We also plot in this figure the relative change in the unit cell volume given by $\Delta V / V = 2 \left(\Delta a / a\right) + \left(\Delta c / c\right) + \left[1 + \left(\Delta c / c\right)\right] \left(\Delta a / a\right)^2 + 2 \left(\Delta a / a\right) \left(\Delta c / c\right)$. The $c$ lattice parameter, as expected, increases monotonically with temperature. The AFM transition has a clear effect on this parameter which shows a kink at $T_N$. The most striking feature of this plot is the anomalous behavior of the $a$ lattice parameter which increases with decreasing temperature and shows absolutely no feature at $T_N$. Despite the anomalous $a$ axis behavior, the total volume of the lattice still increases with increasing temperature. It is also worth mentioning that no change was observed on lattice parameters when a 3~T magnetic field is applied.

%%%%%%%%%%%%%%%%%%%%%%%%%%%%%%%%%%%%%%%%%%%%%%%%%%%%%%%%%%%%%%%%%%%%%%%%%%%%%%%
%
% Discussion
%
\section{Discussion}

As a general rule, a crystalline lattice will contract upon cooling the sample. At a magnetic transition further modifications of the crystalline parameters are induced by the magnetic energy change. Indeed, following the classical work of Baltensperger and Helman~\cite{Baltensperger1968} one can derive the modification of the lattice parameters induced by the magnetic ordering (at a given temperature $T$ below $T_N$) by the minimization of the free energy (${\cal F}$). Below $T_N$, the entropy term of the free energy is negligible and ${\cal F}$ can be written as the sum of the magnetic (${\cal E}_M$) and the elastic (${\cal E}_e$) energies. A Taylor's expansion of ${\cal F}$ as a function of the lattice parameters $a$ and $c$ yields
%
%\begin{linenomath}
\begin{equation} 
  \begin{split}
  	{\cal F}(a,c) &\simeq {\cal E}(a,c)  = {\cal E}_M(a,c) + {\cal E}_e(a,c) \\[1ex]
         &=  \left[{\cal E} +  \frac{\partial {\cal E}}{\partial a} da + \frac{\partial {\cal E}}{\partial c} dc + \cdots \right]_{ (a_N,c_N)}
  \end{split}
\end{equation}
%\end{linenomath}
%
where $(a_N,c_N)$ are the lattice parameters at $T<T_N$ in the antiferromagnetic phase. Naming $a_0,c_0$ the lattice parameters at the same temperature in an hypothetic paramagnetic phase, one gets
%
%\begin{linenomath}
\begin{equation}
  \begin{array}{ll} 
    \frac{\partial {\cal E}}{\partial a}(a_N,c_N) & = 0 
    = \left[\frac{\partial {\cal E}}{\partial a}  
    + \frac{\partial^2  {\cal E}}{\partial a^2} \Delta a 
    + \frac{\partial^2 {\cal E}}{\partial a \partial c} \Delta c \right]_{(a_0,c_0)} 
    \\[2ex]
    \frac{\partial {\cal E}}{\partial c}(a_N,c_N) & = 0 
    = \left[\frac{\partial {\cal E}}{\partial c}
    + \frac{\partial^2 {\cal E}}{\partial a \partial c} \Delta a
    + \frac{\partial^2 {\cal E}}{\partial c^2} \Delta c \right]_{(a_0,c_0)}
  \end{array} %\right. 
\end{equation}
%\end{linenomath}
%
and thus 
%
%\begin{linenomath}
\begin{equation}
	\begin{split}
    \Delta a &= a_N-a_0 = \left.
    \frac{ \frac{\partial {\cal E}}{\partial a} 
    \frac{\partial^2 {\cal E}}{\partial c^2} 
    - \frac{\partial {\cal E}}{\partial c} 
    \frac{\partial^2 {\cal E}}{\partial a \partial c}}
    {\left(\frac{\partial^2 {\cal E}}{\partial a \partial c}\right)^2 
    - \frac{\partial^2 {\cal E}}{\partial a^2}\frac{\partial^2 {\cal E}}{\partial c^2}  }
    \right|_{(a_0,c_0)} 
    \\[2ex]
    \Delta c &= c_N-c_0 = \left. 
    \frac{ \frac{\partial {\cal E}}{\partial c} \frac{\partial^2 {\cal E}}{\partial a^2}
    - \frac{\partial {\cal E}}{\partial a} \frac{\partial^2 {\cal E}}{\partial a \partial c}}
    { \left( \frac{\partial^2 {\cal E}}{\partial a \partial c}\right)^2 
    - \frac{\partial^2 {\cal E}}{\partial a^2}\frac{\partial^2 {\cal E}}{\partial c^2}  }
    \right|_{(a_0,c_0)} 
	\end{split}
	\label{eqdPar}
\end{equation}
%\end{linenomath}
%
The elastic energy per unit cell can be estimated following the classical way of Ref.~\onlinecite{Baltensperger1968}:
%
%\begin{linenomath}
\begin{equation}
{\cal E}_e = \frac{V}{2\kappa}\left(\frac{\Delta V}{V}\right)^2,
\end{equation}
%\end{linenomath}
%
where $V$ is the unit cell volume and $\kappa$ the compressibility coefficient. 

\begin{figure}
  \includegraphics[width=8cm]{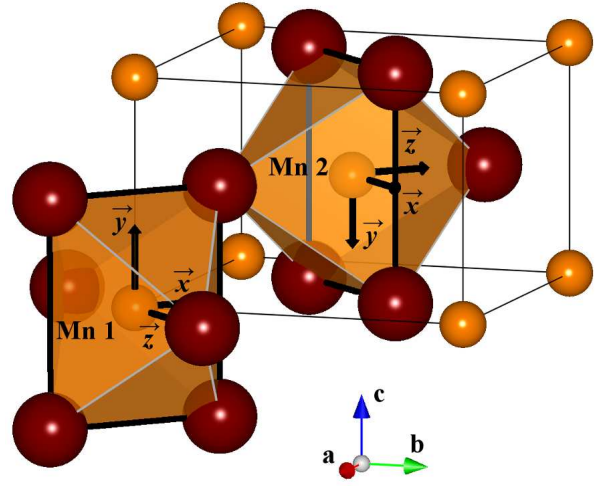}  
  \caption{(color online) Structure of MnF$_2$ highlighting the two possible orientations for the MnF$_6$ coordination octahedra. The thick dark lines connect the in-plane F atoms. The local quantification axes orientations for Mn$_1$ and Mn$_2$ atoms are shown as black arrows. The lattice coordinates framework is defined by the arrows labeled $a$, $b$ and $c$.}
  \label{abinit}
\end{figure}

Figure~\ref{abinit} and Tab.~\ref{tab:axes} show the definitions for the local quantification axes ($x$, $y$ and $z$) of each Mn atom in the unit cell framework. The local Mn $(x,y)$ plane is defined by the in-plane (equatorial) fluorine atoms of the distorted octahedra (shown as the thick line in Fig.~\ref{abinit}), which have the strongest metal-ligand interactions. Note that the two manganese atoms of the unit cell have different local quantification axes. Using this definition, the magnetic energy can be expressed as 
%
%\begin{linenomath}
\begin{equation}
  {\cal E}_M =  - 8 J \langle \vec S_{\rm Mn_1}\cdot\vec S_{\rm Mn_2}\rangle.
  \label{eqMagE}
\end{equation}
%\end{linenomath}
% 
It is important to notice that in the case of MnF$_2$, where the interactions are antiferromagnetic, both $J$ and $\langle \vec S_{\rm Mn_1}\cdot\vec S_{\rm Mn_2}\rangle$ are negative. $J$ can be decomposed into its main superexchange coupling. Using the local quantification axes defined in Fig.~\ref{abinit} and Tab.~\ref{tab:axes}, we determined the superexchange coupling paths and their associated perturbative contributions pictured in Tab.~\ref{tab:coupl}. The sum of these terms leads to: 
%
%\begin{linenomath}
\begin{multline}
  J = -\exp\left[-2 \lambda (d_a + d_e)\right] \times \\
       \left(A_0 + A_2 \, cos^2\alpha + A_4 \, cos^4\alpha + A_6 \, cos^6\alpha\right).
\label{eqJ}
\end{multline}
%\end{linenomath}
%
$d_a$ and $d_e$ represent the apical and equatorial Mn--F distances. $\alpha$ is the angle Mn$_1$--F--Mn$_2$. Although the absolute values of the $A_i$ coefficients are difficult to determine, we can estimate their relative values. We found $A_4 \sim -A_6 \sim 10^2 A_2 \sim 10^3 A_0$. Therefore the last two terms in Eq.\ref{eqJ} dominate the value of $J$.
\begin{table}
  \caption{Local quantification axes of the two Mn atoms in the unit cell. 
    The unit cell framework is expressed in terms of its main $\bf a$, $\bf b$
    and $\bf c$ directions, whereas $\bf x$, $\bf y$ and $\bf z$ refer to the
    local Mn quantification axes.\label{tab:axes}}
\begin{ruledtabular}
\begin{tabular}{ccc} 
	& \multicolumn{2}{l}{Direction in the unit cell framework} \\
  Local axes       & Mn$_1\quad (0,0,0)$ & Mn$_2\quad (1/2,1/2,1/2)$ \\ \hline
   $\vec x$        & $(-\vec a+\vec b)/\sqrt{2}$ & $(\vec a+\vec b)/\sqrt{2}$ \\
   $\vec y$        & $\vec c$                    & $-\vec c$ \\
   $\vec z$        & $(\vec a+\vec b)/\sqrt{2}$  & $(-\vec a+\vec b)/\sqrt{2}$ \\
\end{tabular} 
\end{ruledtabular}
\end{table}
\begin{table}
  \caption{Main AFM coupling paths between adjacent Mn atoms. The Mn $d$ orbitals are written in terms of their local axes (see Tab.~\ref{tab:axes} and Fig.~\ref{abinit}) and the F $p$ orbitals in terms of Mn$_2$ local axes. The vertical and horizontal axes in pictures correspond to the $\vec a+\vec b$ and $-\vec c$ lattice directions, respectively.
    \label{tab:coupl}}
\begin{ruledtabular}
\begin{tabular}{cc} 
   AFM coupling path & $J$ contribution scales as  \\ 
   & ($4^\text{th}$ order perturbation)  \\ \hline
\begin{minipage}{33mm}
  \includegraphics[width=\textwidth]{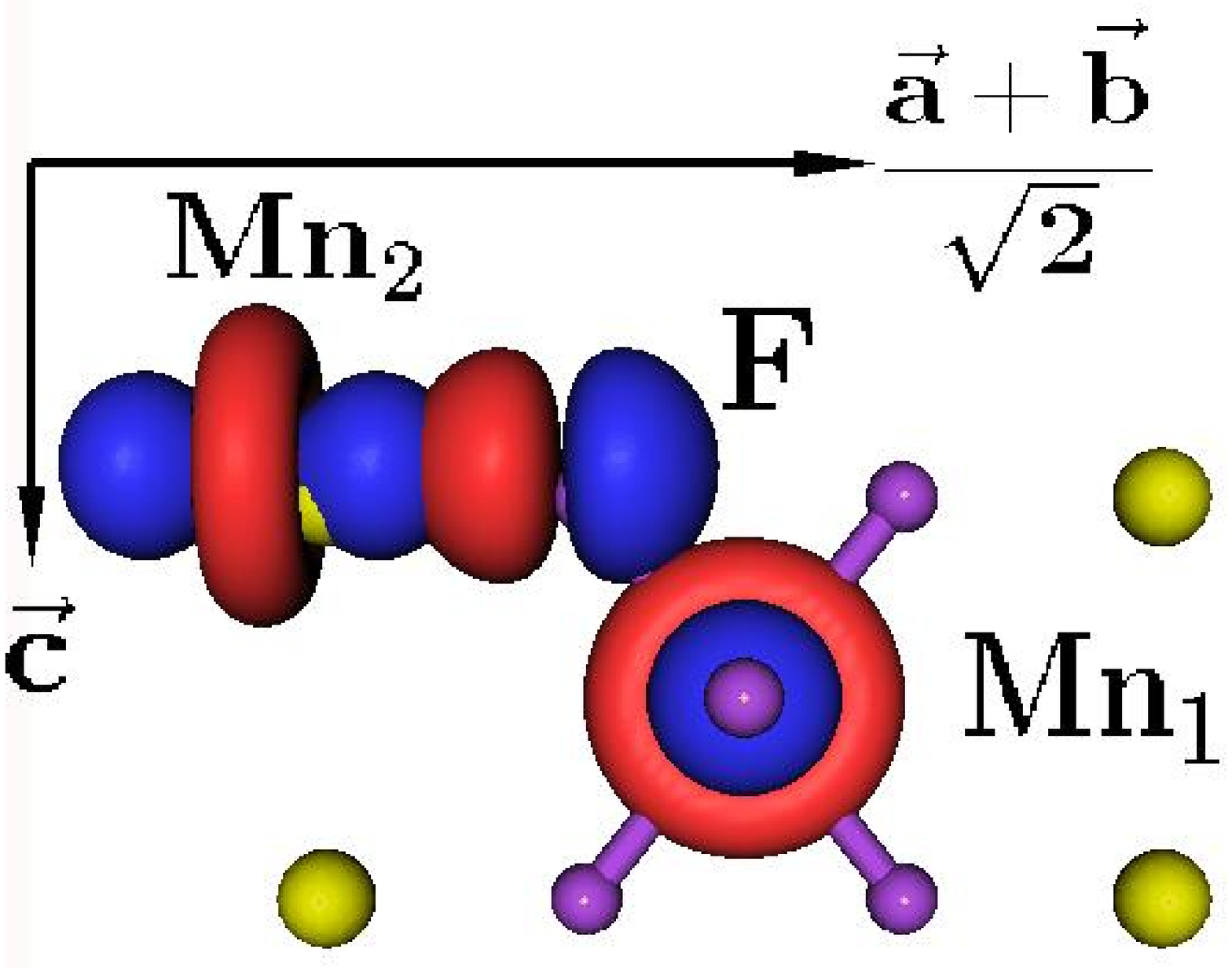}
\end{minipage}
& 
\begin{minipage}{42mm}
  \begin{equation*}
    \begin{split}
      -\langle {\rm Mn}_1\, 3d_{z^2}| {\rm F}\, 2p_x \rangle^2 \times \\
       \langle {\rm F}\, 2p_x | {\rm Mn}_2\, 3d_{z^2} \rangle^2 
    \end{split}
  \end{equation*}
\end{minipage}\\
\begin{minipage}{33mm}
  \includegraphics[width=\textwidth]{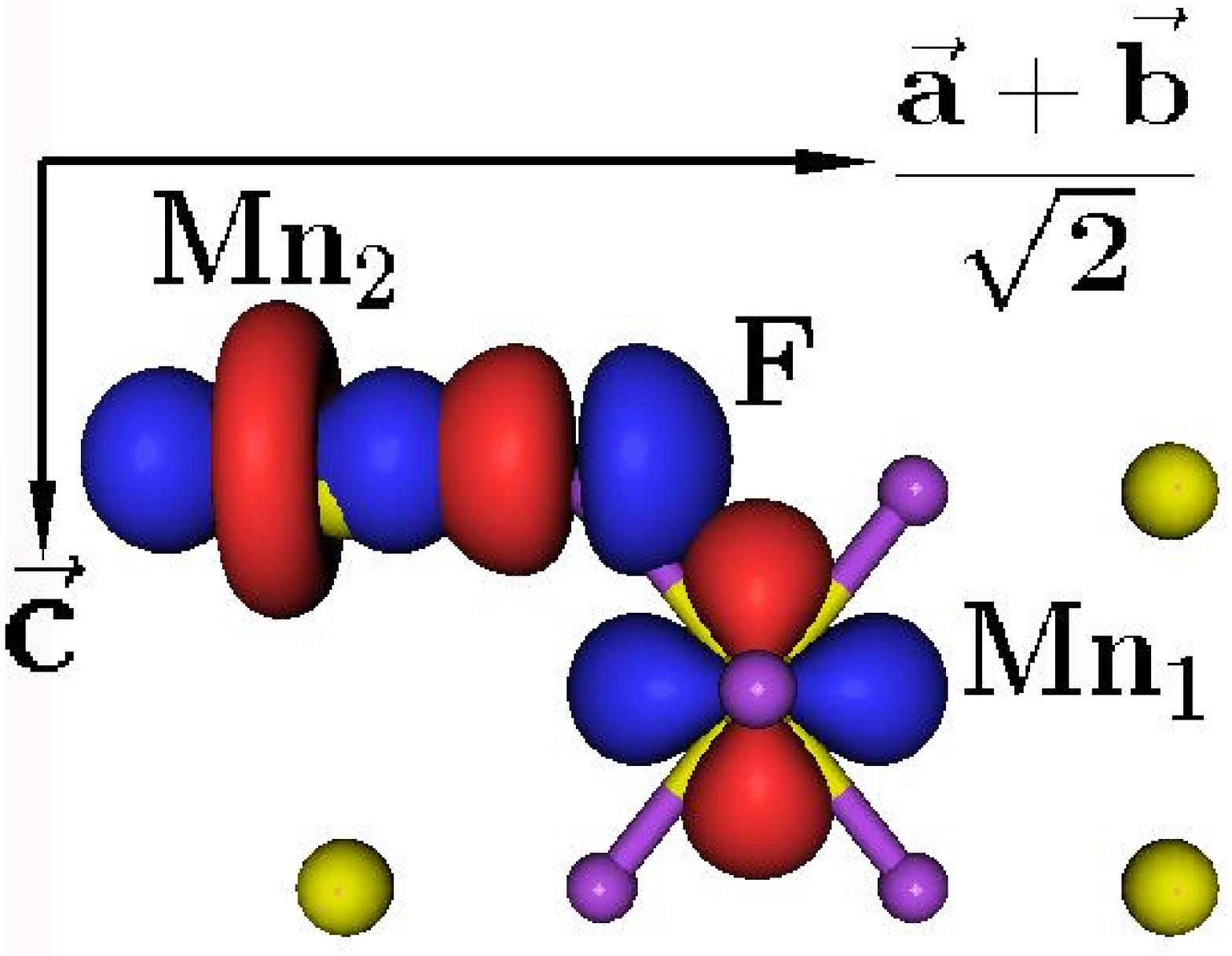}
\end{minipage}
&
\begin{minipage}{42mm}
  \begin{equation*}
    \begin{split}
      -\langle {\rm Mn}_1\, 3d_{z^2} | {\rm F}\, p_x \rangle^2 \times \\
       \langle {\rm F}\, p_x | {\rm Mn}_2\, 3d_{x^2-y^2}\rangle^2
    \end{split}
  \end{equation*}
\end{minipage}\\
\begin{minipage}{33mm}
  \includegraphics[width=\textwidth]{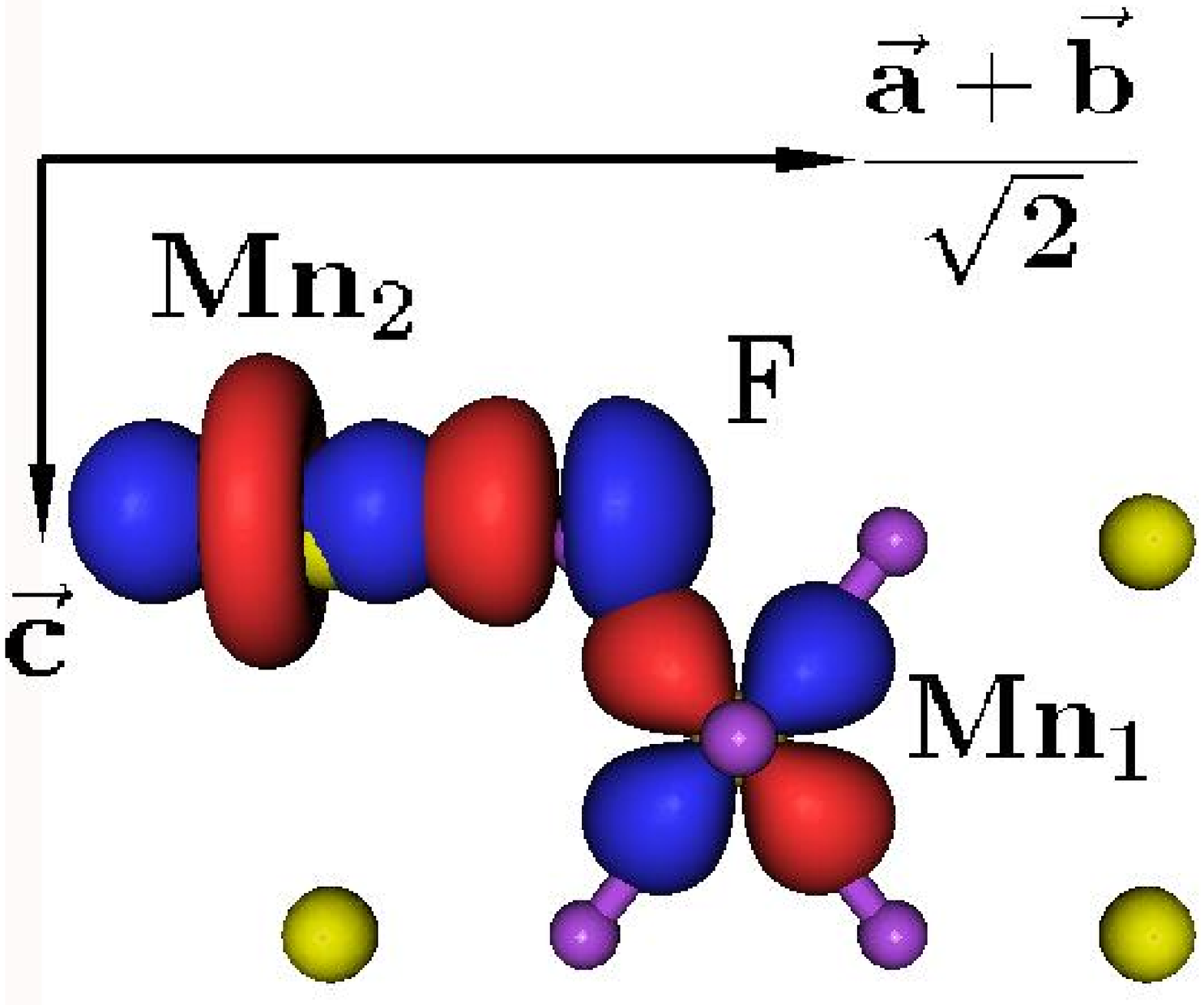}
\end{minipage}
&
\begin{minipage}{42mm}
  \begin{equation*}
    \begin{split}
      -\langle {\rm Mn}_1\, 3d_{z^2}| {\rm F}\, p_x \rangle^2 \times \\
       \langle {\rm F}\, p_x | {\rm Mn}_2\, 3d_{xy}  \rangle^2
    \end{split}
  \end{equation*}
\end{minipage}\\
\begin{minipage}{33mm}
  \includegraphics[width=\textwidth]{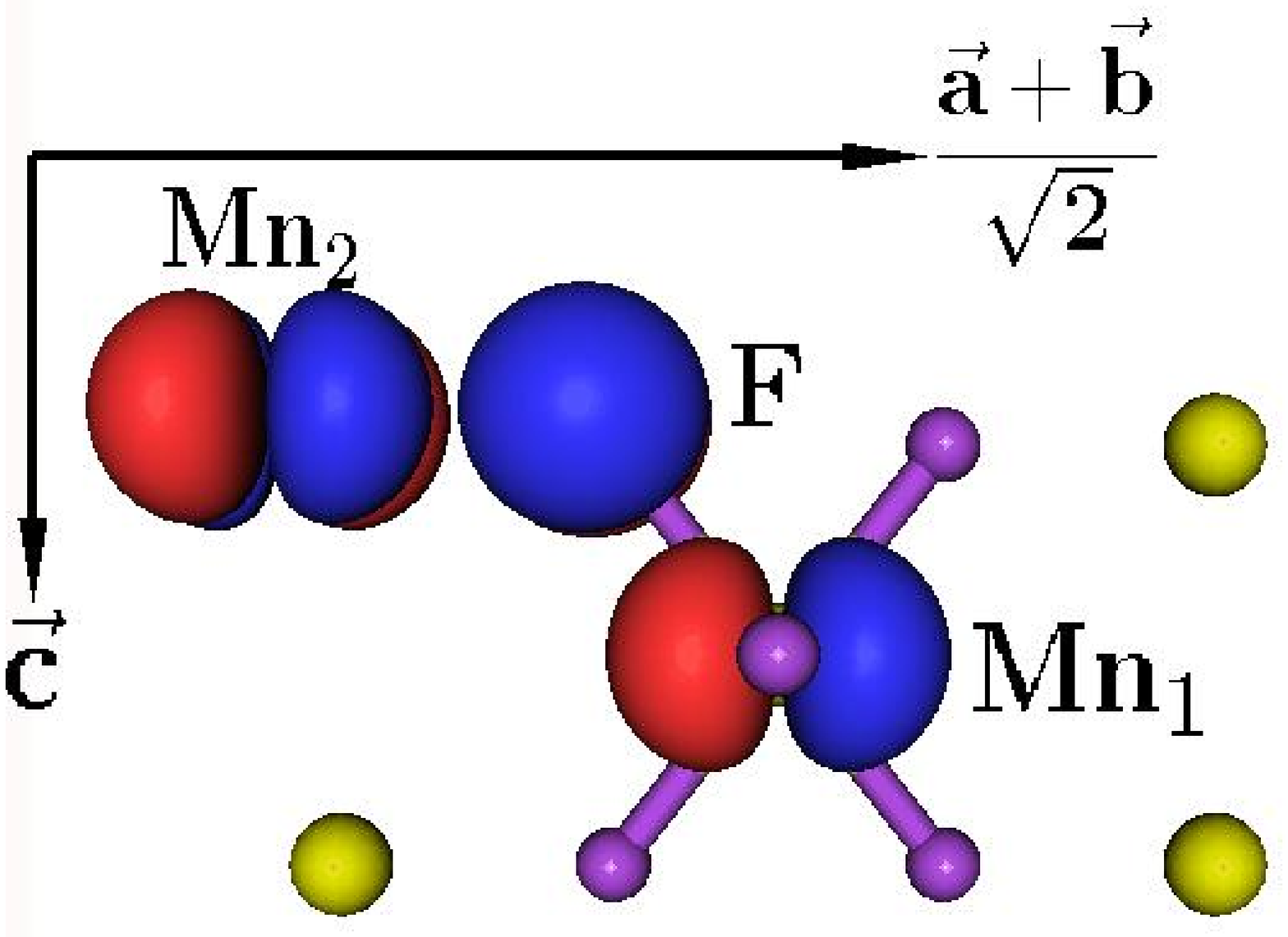}
\end{minipage}
&
\begin{minipage}{42mm}
  \begin{equation*}
    \begin{split}
      -\langle {\rm Mn}_1\, 3d_{xz}| {\rm F}\, 2p_z \rangle^2 \times \\
       \langle {\rm F}\, 2p_z | {\rm Mn}_2\, 3d_{xz}\rangle^2
    \end{split}
  \end{equation*}
\end{minipage}\\
\begin{minipage}{33mm}
  \includegraphics[width=\textwidth]{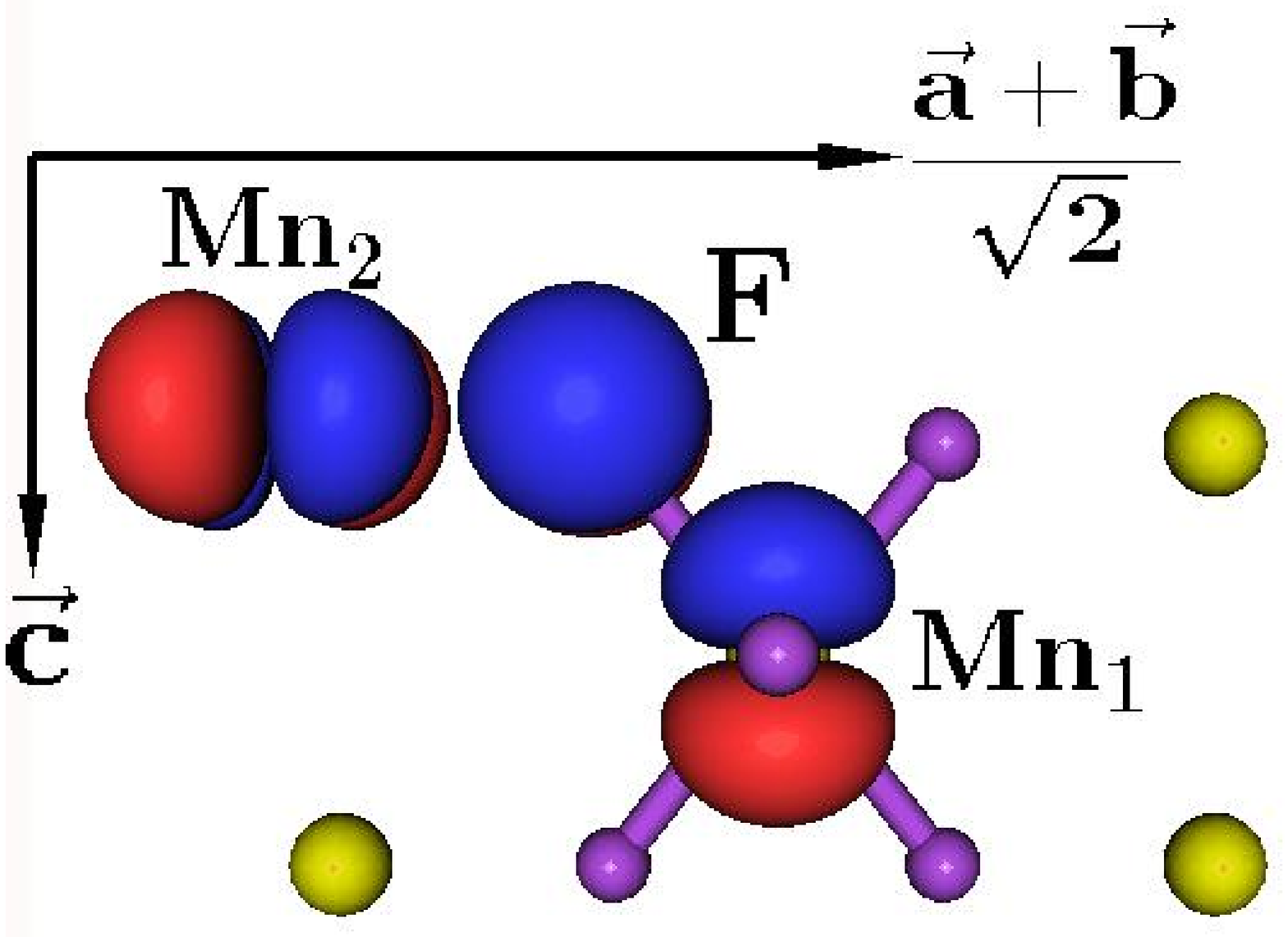}
\end{minipage}
&
\begin{minipage}{42mm}
  \begin{equation*}
    \begin{split}
      -\langle {\rm Mn}_1\, 3d_{xz}| {\rm F}\, 2p_z \rangle^2 \times \\
       \langle {\rm F}\, 2p_z | {\rm Mn}_2\, 3d_{yz}\rangle^2
    \end{split}
  \end{equation*}
\end{minipage}\\
\begin{minipage}{33mm}
  \includegraphics[width=\textwidth]{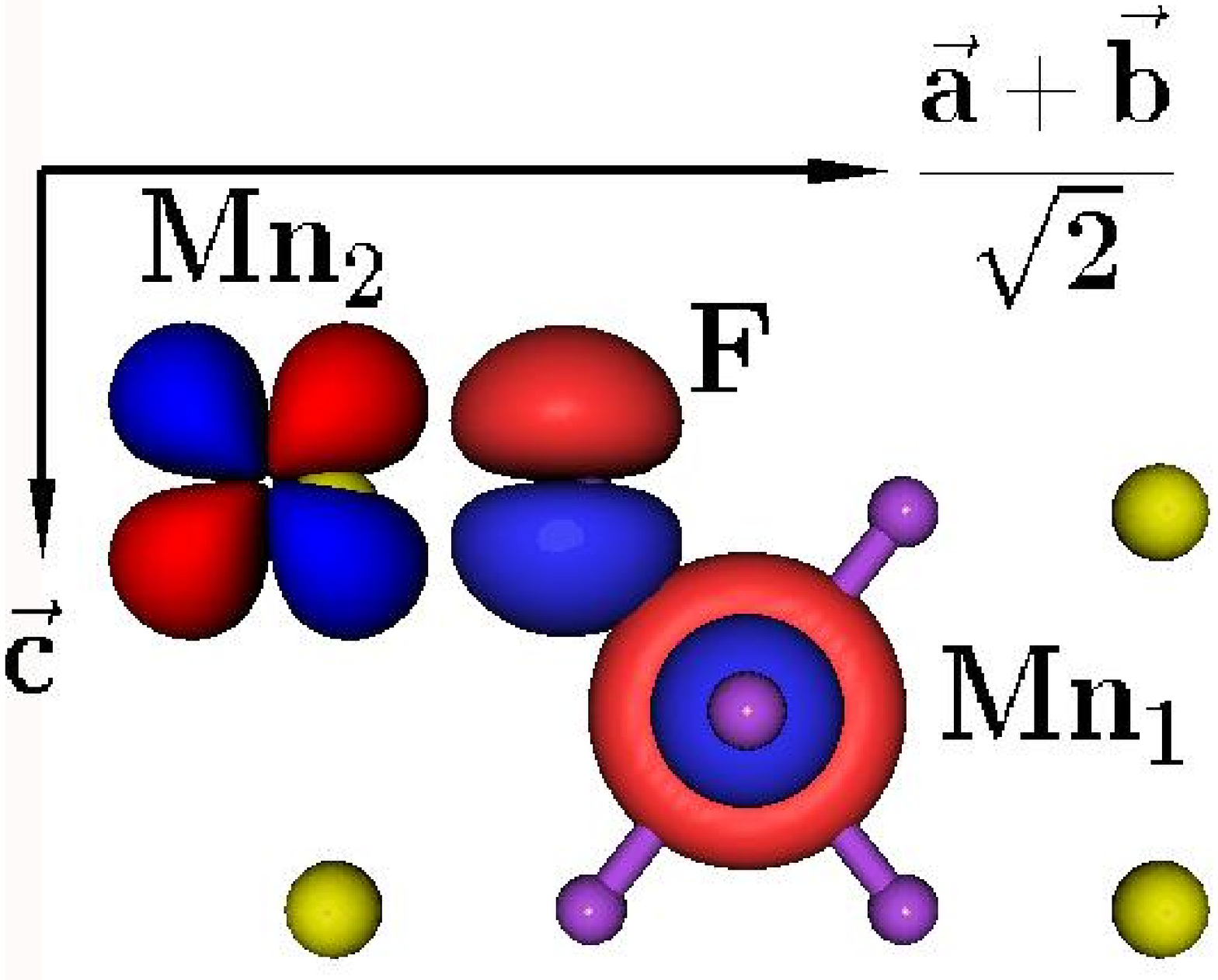}
\end{minipage}
&
\begin{minipage}{42mm}
  \begin{equation*}
    \begin{split}
      -\langle {\rm Mn}_1\, 3d_{yz}| {\rm F}\, 2p_y \rangle^2 \times \\
       \langle {\rm F}\, 2p_y | {\rm Mn}_2\, 3d_{z^2}\rangle^2
    \end{split}
  \end{equation*}
\end{minipage}\\
\begin{minipage}{33mm}
  \includegraphics[width=\textwidth]{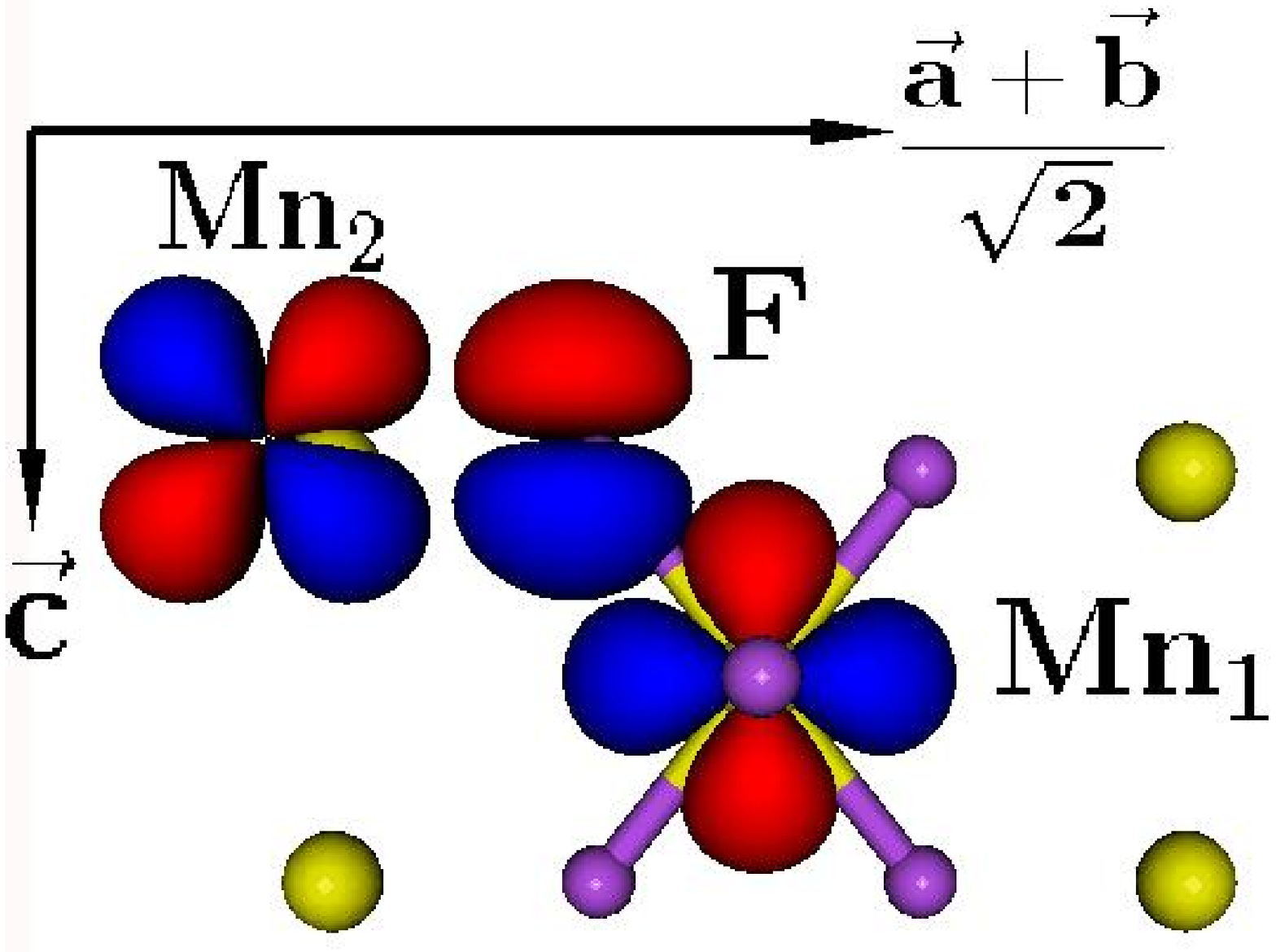}
\end{minipage}
&
\begin{minipage}{42mm}
  \begin{equation*}
    \begin{split}
      -\langle {\rm Mn}_1\, 3d_{yz}| {\rm F}\, 2p_y \rangle^2 \times \\
       \langle {\rm F}\, 2p_y | {\rm Mn}_2\, 3d_{x^2-y^2}\rangle^2  
    \end{split}
  \end{equation*}
\end{minipage}\\
\begin{minipage}{33mm}
  \includegraphics[width=\textwidth]{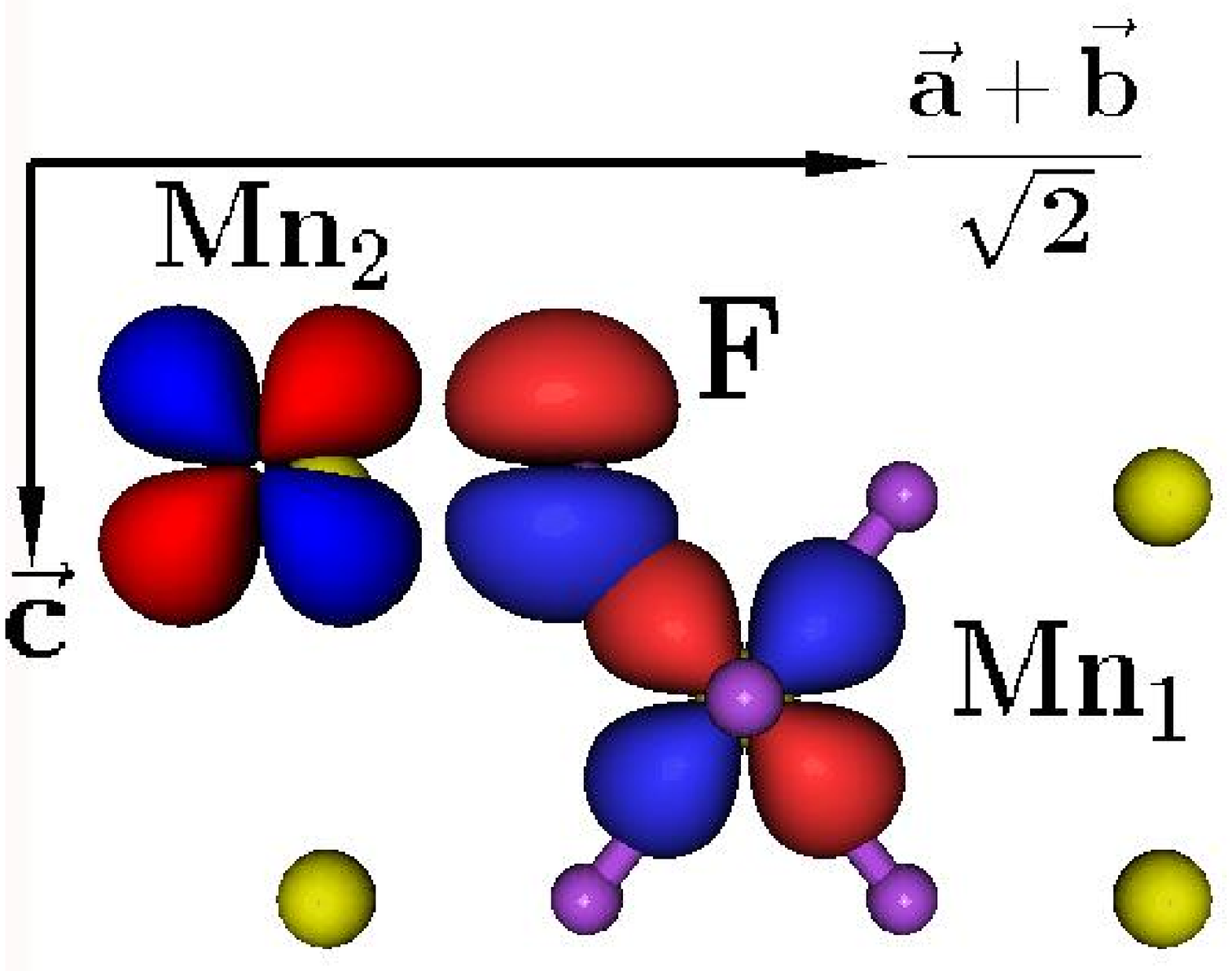}
\end{minipage}
&
\begin{minipage}{42mm}
  \begin{equation*}
    \begin{split}
      -\langle {\rm Mn}_1\, 3d_{yz}| {\rm F}\, 2p_y \rangle^2 \times \\
       \langle {\rm F}\, 2p_y | {\rm Mn}_2\, 3d_{xy}\rangle^2
    \end{split}
  \end{equation*}
\end{minipage}\\
\end{tabular}
\end{ruledtabular}
\end{table}

We can use the above equations to obtain a rough evaluation of the derivatives in Eq.~\ref{eqdPar}. We find that the first derivatives are of the same sign and order of magnitude. This is also the case for $\partial^2{\cal E} / \partial a \partial c$ and $\partial^2{\cal E} / \partial c^2$. On the contrary, ${\partial^2 {\cal E}}/{\partial a^2}$ is nearly an order of magnitude larger than the other second derivative and of opposite sign. The reason is that only in the latter term magnetic and elastic contributions do not contribute with opposite signs. These estimations lead to a very weak value for $\Delta a$, while $\Delta c$ is much larger and of opposite sign, in agreement with the x-ray measurements shown in Fig.~\ref{xray}.

\begin{figure}
  \includegraphics[width=8cm]{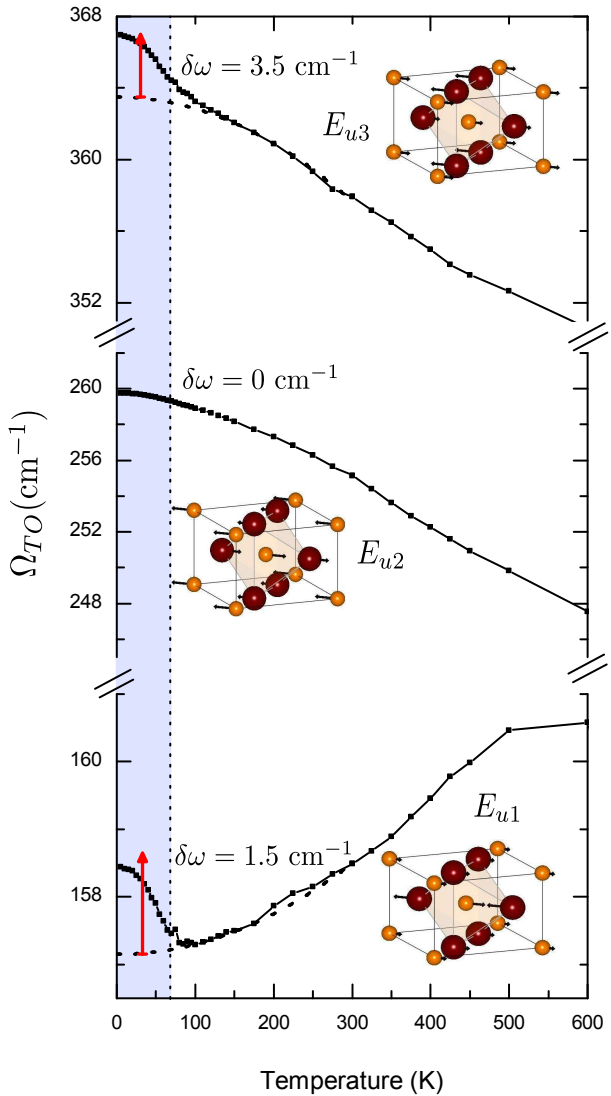}
  \caption{(color online) Evolution of the ab plane phonon frequencies with temperature. The red arrows indicate the \textit{ab-initio} predictions for the change in frequencies (numerically indicated as the $\delta \omega$ values) upon formation of an antiferromagnetic order. The shaded area is the antiferromagnetic phase. For each vibration mode, we represent the atomic motions of manganese and fluorine ions. The dotted lines are guides to the eye showing the high temperature extrapolations of the phonons behavior in the absence of magnetic ordering.}
  \label{freq_a}
\end{figure}

\begin{figure}
  \includegraphics[width=8cm]{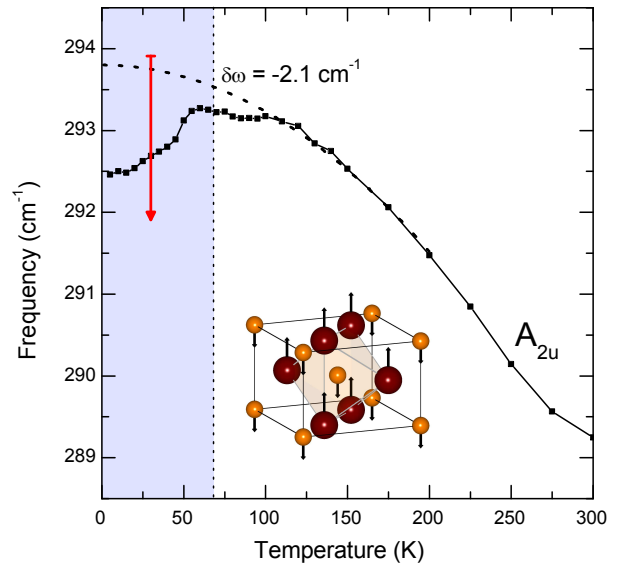}
  \caption{(color online) Evolution of the $c$ axis phonon frequency with temperature. The red arrow indicates the value of the change in frequency when the antiferromagnetic order appears, as predicted by \textit{ab-initio} calculations. The shaded area is the antiferromagnetic phase. The atomic motions of manganese and fluorine ions are given for this vibration mode. The dotted line is a guide to the eye showing the high temperature extrapolation of the phonon behavior in the absence of magnetic ordering.}
  \label{freq_c}
\end{figure}

As the atoms get closer to each other, the interaction forces between them get stronger and, in the harmonic approximation, so does the spring constant ($\kappa$) between ions. As phonon frequencies follow $\sim \sqrt{\kappa / \mu}$ ($\mu$ being a reduced mass), in a first approximation we expect them to follow the thermal evolution of the unit cell volume. Figure~\ref{xray} shows that the volume of the MnF$_2$ lattice increases with temperature. Therefore, we expect phonon frequencies to decrease with increasing temperature. Figures~\ref{freq_a} and \ref{freq_c} show the thermal evolution of phonon frequencies for MnF$_2$. Phonons $E_{u2}$ and $E_{u3}$ show the expected frequency softening with increasing temperature over the whole measured range. Phonon $A_{2u}$ also behaves conventionally, but only above $T_N$.

Phonon $E_{u1}$, on the other hand, has an anomalous thermal evolution as its frequency softens with decreasing temperature. This behavior is analogous to that reported for the $B_{1g}$ mode seen by Raman spectroscopy in MnF$_2$ (Ref.~\onlinecite{LOCKWOOD1988}) and other rutile type fluorides such as ZnF$_2$,\cite{Giordano1988} FeF$_2$,\cite{LOCKWOOD1983} NiF$_2$,\cite{LOCKWOOD2002} and MgF$_2$.\cite{PERAKIS1999} This mode seems to be more influenced by the $a$ parameter rather than the whole unit cell volume. The unconventional lattice parameters thermal evolution and this anomalous phonon behavior are strong indications of large lattice instabilities.

The existence of these lattice instabilities is emphasized by Raman scattering under pressure. In ZnF$_2$ the softening of the $B_{1g}$ Raman mode with decreasing temperature is also observed by applying pressure at room temperature. This pressure induced phonon softening is a precursor of a phase transition from rutile to a CaCl$_2$ structure at 4.5 GPa.\cite{PERAKIS2005} Brillouin scattering measurements\cite{YAMAGUCHI1992} showed that MnF$_2$ also undergoes a phase transition at 1.49 GPa, whose nature was inconclusive. Since the transition in ZnF$_2$ is accompanied by an orthorhombic distortion of the lattice, we can interpret the $E_{u1}$ mode softening in MnF$_2$ as a precursor of a (incipient) phase transition. Further insight on this transition is given by measurements of the elastic properties in rutiles. Melcher\cite{Melcher1970} showed that the $c_{44}$ elastic constant in MnF$_2$ has one anomaly at $T_N$ and another, unrelated to magnetism, at lower temperatures. Rimai\cite{Rimai1977} observed a similar low temperature accident in the non magnetic rutile ZnF$_2$ and concluded that it was representative of a structural instability compatible with ferroelectricity. Hence, although only new measurements will be able to settle the issue, it is reasonable to expect that the high pressure transition in MnF$_2$ is of (incipient) ferroelectric character.

Figures~\ref{freq_a} and \ref{freq_c} show that at the AFM transition, there is a clear renormalization of the phonon spectra. Mode $E_{u3}$ shows an additional increase in its frequency below $T_N$. This is an effect compatible with the magnetostrictive kink observed in the lattice volume, which decreases faster below $T_N$. There is no noticeable change in the behavior of phonon $E_{u2}$. A puzzling behavior is observed for phonons $E_{u1}$ and $A_{2u}$. The phonon softening instability observed in mode $E_{u1}$ stops at $T_N$ and the frequency of this phonon increases notably with the appearance of the AF phase. Conversely, the thermal evolution of mode $A_{2u}$ reverses, and this phonon shows a frequency softening below $T_N$, which is at odds with the lattice volume evolution. As no lattice parameter show a sign change in their thermal behavior at $T_N$, the sign change in the slope of the thermal evolution of these two phonon modes is likely due to effects other than conventional magnetostriction, such as exchange between ions.

To grasp further insight on this phonon behavior, we can look into \textit{ab-initio} results. The paramagnetic phase is not accessible to \textit{ab-initio} calculations. We considered that the average of calculations in the AFM and FM configurations are representative of the paramagnetic phase. The difference in the phonon frequencies between the AFM and ``paramagnetic'' calculations are indicated by the vertical arrows in Figs.~\ref{freq_a} and \ref{freq_c}. It is remarkable that not only do \textit{ab-initio} calculations predict the correct magnitude for the phonon frequency shifts, but also they find the proper sign for these changes. It is important to remark that \textit{ab-initio} calculations are done at 0~K. Hence, its predictions for phonon changes are not due to lattice modifications but are representative of a direct coupling of the phonons to the magnetic ordering, without the need of mediation by elastic lattice distortions.

\textit{Ab-initio} calculations also yield the normal mode eigenvectors. One can thus analyze the effect of the displacements associated with these phonon modes on the magnetic coupling constant. The infrared active normal modes are shown in Figs.~\ref{freq_a} and \ref{freq_c}. The thermal evolution of all phonons in the paramagnetic phase, including the anomalous softening when lowering the temperature of the $E_{u1}$ mode is probably dominated by the elastic energy. Indeed, the same behavior is observed in the Raman spectra of the $B_{1g}$ mode of MgF$_2$ (Ref.~\onlinecite{PERAKIS1999}) and the infrared response of the $A_{2u}$ mode of ZnF$_2$ (Ref.~\onlinecite{Giordano1987}), both non magnetic fluorides, isostructural to MnF$_2$.

In order to understand the sign change in the thermal evolution of phonons $E_{u1}$ and $A_{2u}$ below $T_N$, we have to look at the free energy ${\cal F} = {\cal E} -TS$. In the AFM phase, the dominant term in the changes of the free energy of the system is the magnetic energy ${\cal E}_M$, given by Eq.~\ref{eqMagE}. Thus, minimization of ${\cal F}$ goes through the minimization of ${\cal E}_M$. Let us look into the contribution coming from phonons $E_{u1}$ and $A_{2u}$ to this term. The $E_{u1}$ mode corresponds to a large displacement in the $\vec a+\vec b$ and $\vec a-\vec b$ directions of the F atoms bridging two MnF$_6$ octahedra. Let us define $J_{\vec 0}$ as the magnetic coupling between the Mn$_2$ atom and the Mn$_1$ atoms located at the crystallographic positions $(0,0,0)$ and $(0,0,1)$; $J_{\vec a+\vec b}$ as the coupling between Mn$_2$ and the Mn$_1$ atoms located at $(1,1,0)$ and $(1,1,1)$; and $J_{\vec a}$ as the coupling between Mn$_2$ and the Mn$_1$ atoms located at $(1,0,0)$, $(1,0,1)$, $(0,1,0)$ and $(0,1,1)$. Under a $E_{u1}$ phonon displacement of amplitude $u$, the magnetic energy is modified as follows
%
%\begin{linenomath}
\begin{multline} 
  {\cal E}_M = -\left[2 \,J_{\vec 0}(u)  + 2 J_{\vec a+\vec b}(u) 
                  +4 J_{\vec a}(u)\right] 
                  \langle \vec S_{\rm Mn_1} \cdot 
                              \vec S_{\rm Mn_2}\rangle \\
                = -\left[8 J(0) + 2 \left.\frac{\partial^2 J_{\vec 0}}
                {\partial u^2}\right|_{u=0} u^2 \right]
                \langle \vec S_{\rm Mn_1} \cdot \vec S_{\rm Mn_2}\rangle. 
  \label{eqEMEu1}
\end{multline}
%\end{linenomath}
%
$J(0)$ is the value of the magnetic exchange with the atoms in their equilibrium position. We should stress, again, that both $J(0)$ and the spin correlation terms are negative because of the AFM interaction. Using the perturbative expressions of Tab.~\ref{tab:coupl} one can show that $J_{\vec a}$ does not depend on the $E_{u1}$ displacement $u$, that $\partial J_{\vec 0}/\partial u = - \partial J_{\vec a+\vec b}/\partial u$, and that $\partial^2 J_{\vec 0}/\partial u^2 = \partial^2 J_{\vec a+\vec b}/\partial u^2 > 0$. In the AFM phase the magnetic energy must be minimized, and thus, the modulus of the first factor of Eq.~\ref{eqEMEu1} must increase. As the second derivative and $J(0)$ have opposite signs, increasing the modulus of the first factor implies that the oscillation amplitude $u$ must decrease. In a equivalent view, one can note that the second derivative term acts as a correction to the full harmonic potential of the phonon. For phonon $E_{u1}$ it introduces a positive correction to the effective phonon spring constant, \textit{i.e.} the phonon hardens, as observed in Fig.~\ref{freq_a}.

A similar analysis also explains why the $A_{2u}$ mode softens when the temperature decreases in the AFM phase. The $A_{2u}$ mode corresponds to displacements with opposite signs of the Mn and F atoms along the $c$ direction (see Fig~\ref{freq_c}). We define $J_{\vec 0}$ as the magnetic coupling between the Mn$_2$ atom and the Mn$_1$ atoms located at $(0,0,0)$, $(1,1,0)$, $(1,0,1)$, $(0,1,1)$; and $J_{\vec c}$ as the coupling between the Mn$_2$ atom and the Mn$_1$ atoms located at $(0,0,1)$, $(1,1,1)$, $(1,0,0)$, $(0,1,0)$. Under a $A_{2u}$ displacement of amplitude $v$, the magnetic energy is modified as 
%
%\begin{linenomath}
\begin{multline}   
  {\cal E}_M = -\left[4 J_{\vec 0}(v)+4 J_{\vec c}(v)\right]
               \langle \vec S_{\rm Mn_1} \cdot \vec S_{\rm Mn_2}\rangle \\
               -\left[8J(0) + 4\left.\frac{\partial^2 J_{\vec 0}}
               {\partial v^2}\right|_{v=0} v^2 \right]
               \langle \vec S_{\rm Mn_1} \cdot \vec S_{\rm Mn_2}\rangle.
\end{multline}
%\end{linenomath}
%
Using again a rough evaluation of the perturbative expression of $J$ one finds that $\partial J_{\vec 0}/\partial v = - \partial J_{\vec c}/\partial v$, and $\partial^2 J_{\vec 0}/\partial v^2 = \partial^2 J_{\vec c}/\partial v^2 < 0$. Because the second derivative and $J(0)$ here have the same sign, an increase of the displacement amplitude $v$ diminishes the value of ${\cal E}_M$. In the full harmonic potential perspective, here the correction to the effective phonon spring constant is negative, leading to the phonon softening,  in agreement with Fig.~\ref{freq_c}.

\begin{figure}
  \includegraphics[width=8cm]{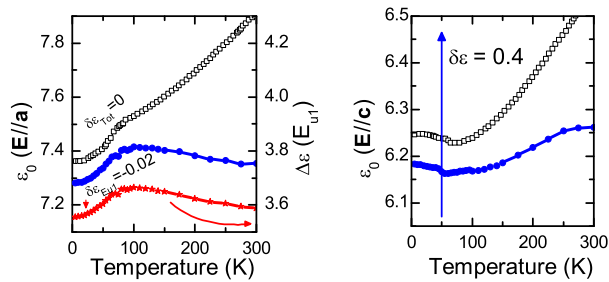}  
  \caption{(color online) The left panel shows the $ab$ plane dielectric constant $\varepsilon(0)$ measured utilizing dielectric techniques by Seehra\protect\cite{SEEHRA1986} (open squares) and zero frequency optical extrapolation (solid circles). The right panel shows the same quantities for the $c$ axis. For clarity, the dielectric measurement was shifted by -1. The blue arrows are the \textit{ab-initio} predictions for changes $\delta\varepsilon$ in $\varepsilon(0)$ upon appearance of the antiferromagnetic order. The right-hand scale in the left panel shows the $E_{u1}$ contribution to the dielectric constant (solid stars) and the red arrow is the \textit{ab-initio} prediction for the change in $\Delta\varepsilon \left(E_{u1}\right)$. Note that right and left scales in this panel span the same range in values of $\varepsilon$, indicating that all the in-plane changes come from the $E_{u1}$ mode.}
  \label{eps0}
\end{figure}

Our infrared data allow us to link the magnetic ordering to changes in the dielectric constant of MnF$_2$. The optical dielectric constant is given by $\varepsilon(0) = \varepsilon_\infty + \sum_k \Delta\varepsilon_k$. In the absence of any microwave excitations, it should be equal to the \textit{static} value obtained from electrical measurements. Our reflectivity data do not show any extra excitation below the phonons down to 10~\icm\ (300 GHz). Figure~\ref{eps0} compares the static dielectric constants, measured by Seehra\cite{SEEHRA1986} at $10$~kHz, with the zero frequency limit of our optical data. The left panel shows the $ab$-plane dielectric constant, and the right panel the $c$ axis response. The solid squares are the ``static'' values from Seehra\cite{SEEHRA1986} (vertically shifted by -1, for clarity) and the open circles are the optical dielectric constants. On the left panel, we also show the thermal variation of the oscillator strength for the lowest energy $E_{u1}$ mode (solid triangles, right-hand scale). Note that both scales in this figure cover the same variation in values of $\varepsilon$. It is then clear that all the temperature variations in the optical dielectric constant come from the $E_{u1}$ phonon. This figure also shows that, within 15\%, the static dielectric constant is build from phonon and higher frequency excitations. The static and optical dielectric constants agree very well below $T_N$. In the paramagnetic phase, there is a growing difference between these two values. Along the $c$-axis both values increase with temperature but the static dielectric constant increases faster. In the $ab$-plane the difference is also qualitative. The static dielectric constant increases with temperature, whereas the optical value decreases. These discrepancies may be due to a difference in the quality of the sample, but are more likely linked to ionic conduction contributions to the dielectric constant in the radiofrequency range, a common effect on fluorides.\cite{Ure1957,SAMARA1979} Nevertheless, in both polarizations, $\varepsilon(0)$, measured by both techniques, shows a kink at $T_N$. 

We can estimate the \textit{ab-initio} zero frequency dielectric constant from the calculated oscillator strengths. We made the same comparison between a ``paramagnetic'' and AFM configurations, as we did for the phonon frequencies. The predicted changes are shown as the vertical arrows in Fig.~\ref{eps0}. The predicted \textit{ab-initio} jump upon magnetic ordering is of the same order of magnitude as the values measured. More important, we find the correct sign for the changes in both cases. The AFM order tends to decrease the dielectric constant in the $ab$-plane while it is increased along the $c$-axis.

The panorama in MnF$_2$ points towards instabilities that favor a ferroelectric order. Ferroelectricity is not a common property of fluorides. However, several compounds (ZnF$_2$, NiF$_2$ and FeF$_2$) show lattice instabilities that indicate the appearance of an incipient ferroelectric order. Our x-ray data indicate that lattice instabilities are indeed present in MnF$_2$. Infrared measurements show that the $c$-axis phonon softens upon formation of the antiferromagnetic order. This softening goes together with an increase of the dielectric constant along this direction. These effects are indicative of a possible incipient ferroelectric behavior in MnF$_2$. In this scenario the differences between static and phononic dielectric constants observed in Fig.~\ref{eps0} above 100 K could also be related to zero frequency ferroelectric fluctuations. However, dielectric constant measurements at high frequencies as well as phonon dynamic measurements under pressure and/or magnetic field are necessary to confirm (or refuse) this picture. In any case, understanding the phonon changes induced by the magnetic ordering in MnF$_2$ should bring new insights on the magneto-electric coupling in multiferroic materials.

%%%%%%%%%%%%%%%%%%%%%%%%%%%%%%%%%%%%%%%%%%%%%%%%%%%%%%%%%%%%%%%%%%%%%%%%%%%%%%%
%
% Conclusions
%
\section{Conclusions}

In this paper we presented a detailed temperature dependent infrared study of the phonon renormalization observed at $T_N$ in MnF$_2$. We showed that the phonon temperature dependence and the lattice parameter changes across the antiferromagnetic transition are well reproduced by \textit{ab-initio} calculations, implying that the magnetic order dominates the changes observed at T$_N$. We find that phonons along the $D_4$ axis and the tetragonal $ab$ plane have opposite changes at $T_N$ as predicted by the first principles results. Our results show that the dielectric constant of MnF$_2$ is mostly from phonon origin. Relaxation effects on the $ab$ plane contribute strongly to the dielectric constant in the paramagnetic phase. The large change in the lattice parameters with temperature and the phonon softening in the antiferromagnetic phase suggest that MnF$_2$ distortions induced by the magnetic order are compatible with the ferroelectric instabilities observed in TiO$_2$ and FeF$_2$ and other fluorides. We suggest that the phase transition observed at 1.49 GPa (Ref.~\onlinecite{YAMAGUCHI1992}) could be of ferroelectric order and so, MnF$_2$ would qualify as an incipient multiferroic.

\section*{Acknowledgment}

We thank Prof. J.-Y. Gesland for providing us with the MnF$_2$ crystal used in this study. RLM acknowledges an invited ``Joliot chair'' at ESPCI and RPSML acknowledges an invited scientist position from FAPEMIG. This work was partially funded by the Brazilian agencies CNPq and FAPEMIG. The collaboration between ESPCI and UFMG was supported by the CNRS PICS 4905. Use of the NSLS was supported by the U.S. Dept. of Energy, Office of Science, Office of Basic Energy Sciences, under Contract No. DE-AC02-98CH10886. Structure diagrams were produced using VESTA software.\cite{VESTA} The \textit{ab-initio} calculations were performed at the IDRIS and CRIHAN computational centers under projects number 1842 (IDRIS) and 2007013 (CRIHAN).

\section*{Appendix}

In this paper, we limited our discussion of \textit{ab-initio} calculations to the four infrared active modes only. For completeness, Table~\ref{tababinit} gives the results of calculations for all modes (infrared, Raman and silent) in both FM and pseudo-AFM configurations. These calculations are compared to the experimental values determined at 5~K for the infrared (this work) and Raman active modes (Ref.~\onlinecite{LOCKWOOD1988}).
\begin{table}[H]
\begin{center}
\caption{\textit{Ab-initio} $\Gamma$ point optical phonons frequencies (in \icm) for the FM and pseudo-AFM phases. The experimental values were taken at 5~K from the present work (IR) and from Ref.~\protect\onlinecite{LOCKWOOD1988} (Raman)}.
\label{tababinit}
\begin{ruledtabular}
\begin{tabular}{cc....}
 & & \multicolumn{2}{c}{\textit{ab-initio}} & \multicolumn{2}{c}{Experimental}
\\
\cline{3-4}\cline{5-6} Symmetry  & Activity  & \multicolumn{1}{c}{FM} &  \multicolumn{1}{c}{AFM}
 & \multicolumn{1}{c}{This work} &
\multicolumn{1}{c}{Ref.~\onlinecite{LOCKWOOD1988}}\\

\hline
$B_{1g}$ & Raman &  96.2 &  97.8 & & 56.1\\
$B_{1u}$ & -  & 136.2 &  136.4& &  \\
$E_u$    &IR & 175.2 &  178.2&158.4 \\  
$A_{2g}$ & -  & 238.9 &  241.1&  \\ 
$E_g$    & Raman & 257.8 &  257.2& &246.7\\  
$E_u$    & IR & 265.5 &  265.4&259.8 \\  
$A_{2u}$ & IR & 315.1 &  311.0&292.5 \\  
$B_{1u}$ & -   & 346.7 &  344.4& \\  
$A_{1g}$ & Raman  & 365.8 &  368.7& & 346.8\\  
$E_u$    & IR & 387.7 &  394.6&367.0 \\  
$B_{2g}$ & Raman & 490.6 &  491.6 & & 480.5\\ 
\end{tabular}
\end{ruledtabular}
\end{center}
\end{table}

%
% The bibliography (BibTeX)
%
\bibliography{mnf2_refs}

\begin{thebibliography}{44}
\expandafter\ifx\csname natexlab\endcsname\relax\def\natexlab#1{#1}\fi
\expandafter\ifx\csname bibnamefont\endcsname\relax
  \def\bibnamefont#1{#1}\fi
\expandafter\ifx\csname bibfnamefont\endcsname\relax
  \def\bibfnamefont#1{#1}\fi
\expandafter\ifx\csname citenamefont\endcsname\relax
  \def\citenamefont#1{#1}\fi
\expandafter\ifx\csname url\endcsname\relax
  \def\url#1{\texttt{#1}}\fi
\expandafter\ifx\csname urlprefix\endcsname\relax\def\urlprefix{URL }\fi
\providecommand{\bibinfo}[2]{#2}
\providecommand{\eprint}[2][]{\url{#2}}

\bibitem[{\citenamefont{Khomskii}(2009)}]{KHOMSKII2009}
\bibinfo{author}{\bibfnamefont{D.}~\bibnamefont{Khomskii}},
  \bibinfo{journal}{Physics} \textbf{\bibinfo{volume}{2}}, \bibinfo{pages}{20}
  (\bibinfo{year}{2009}).

\bibitem[{\citenamefont{Smolenskii and Chupis}(1959)}]{Smolenskii1959}
\bibinfo{author}{\bibfnamefont{G.~A.} \bibnamefont{Smolenskii}}
  \bibnamefont{and} \bibinfo{author}{\bibfnamefont{I.~E.}
  \bibnamefont{Chupis}}, \bibinfo{journal}{Usp. Fiz. Nauk.}
  \textbf{\bibinfo{volume}{137}}, \bibinfo{pages}{415} (\bibinfo{year}{1959}).

\bibitem[{\citenamefont{Kiselev et~al.}(1963)\citenamefont{Kiselev, Oserov, and
  Zhdanov}}]{Kiselev1963}
\bibinfo{author}{\bibfnamefont{S.~V.} \bibnamefont{Kiselev}},
  \bibinfo{author}{\bibfnamefont{R.~P.} \bibnamefont{Oserov}},
  \bibnamefont{and} \bibinfo{author}{\bibfnamefont{G.~S.}
  \bibnamefont{Zhdanov}}, \bibinfo{journal}{Sov. Phys. Dokl.}
  \textbf{\bibinfo{volume}{7}}, \bibinfo{pages}{742} (\bibinfo{year}{1963}).

\bibitem[{\citenamefont{Kimura et~al.}(2003)\citenamefont{Kimura, Goto,
  Shintani, Ishizaka, Arima, and Tokura}}]{Kimura2003}
\bibinfo{author}{\bibfnamefont{T.}~\bibnamefont{Kimura}},
  \bibinfo{author}{\bibfnamefont{T.}~\bibnamefont{Goto}},
  \bibinfo{author}{\bibfnamefont{H.}~\bibnamefont{Shintani}},
  \bibinfo{author}{\bibfnamefont{K.}~\bibnamefont{Ishizaka}},
  \bibinfo{author}{\bibfnamefont{T.}~\bibnamefont{Arima}}, \bibnamefont{and}
  \bibinfo{author}{\bibfnamefont{Y.}~\bibnamefont{Tokura}},
  \bibinfo{journal}{Nature} \textbf{\bibinfo{volume}{426}}, \bibinfo{pages}{55}
  (\bibinfo{year}{2003}).

\bibitem[{\citenamefont{Schmidt et~al.}(2009)\citenamefont{Schmidt, Kant,
  Rudolf, Mayr, Mukhin, Balbashov, Deisenhofer, and Loidl}}]{Schmidt2009}
\bibinfo{author}{\bibfnamefont{M.}~\bibnamefont{Schmidt}},
  \bibinfo{author}{\bibfnamefont{C.}~\bibnamefont{Kant}},
  \bibinfo{author}{\bibfnamefont{T.}~\bibnamefont{Rudolf}},
  \bibinfo{author}{\bibfnamefont{F.}~\bibnamefont{Mayr}},
  \bibinfo{author}{\bibfnamefont{A.~A.} \bibnamefont{Mukhin}},
  \bibinfo{author}{\bibfnamefont{A.~M.} \bibnamefont{Balbashov}},
  \bibinfo{author}{\bibfnamefont{J.}~\bibnamefont{Deisenhofer}},
  \bibnamefont{and} \bibinfo{author}{\bibfnamefont{A.}~\bibnamefont{Loidl}},
  \bibinfo{journal}{Eur. Phys. J. B} \textbf{\bibinfo{volume}{71}},
  \bibinfo{pages}{411} (\bibinfo{year}{2009}).

\bibitem[{\citenamefont{Wakamura and Arai}(1988)}]{Wakamura1988}
\bibinfo{author}{\bibfnamefont{K.}~\bibnamefont{Wakamura}} \bibnamefont{and}
  \bibinfo{author}{\bibfnamefont{T.}~\bibnamefont{Arai}}, \bibinfo{journal}{J.
  Appl. Phys.} \textbf{\bibinfo{volume}{63}}, \bibinfo{pages}{5824}
  (\bibinfo{year}{1988}).

\bibitem[{\citenamefont{Rudolf et~al.}(2007)\citenamefont{Rudolf, Kant, Mayr,
  Hemberger, Tsurkan, and Loidl}}]{Rudolf2007}
\bibinfo{author}{\bibfnamefont{T.}~\bibnamefont{Rudolf}},
  \bibinfo{author}{\bibfnamefont{C.}~\bibnamefont{Kant}},
  \bibinfo{author}{\bibfnamefont{F.}~\bibnamefont{Mayr}},
  \bibinfo{author}{\bibfnamefont{J.}~\bibnamefont{Hemberger}},
  \bibinfo{author}{\bibfnamefont{V.}~\bibnamefont{Tsurkan}}, \bibnamefont{and}
  \bibinfo{author}{\bibfnamefont{A.}~\bibnamefont{Loidl}},
  \bibinfo{journal}{New J. Phys.} \textbf{\bibinfo{volume}{9}},
  \bibinfo{pages}{76} (\bibinfo{year}{2007}).

\bibitem[{\citenamefont{Sushkov et~al.}(2005)\citenamefont{Sushkov,
  Tchernyshyov, Ratcliff~II, Cheong, and Drew}}]{Sushkov2005}
\bibinfo{author}{\bibfnamefont{A.~B.} \bibnamefont{Sushkov}},
  \bibinfo{author}{\bibfnamefont{O.}~\bibnamefont{Tchernyshyov}},
  \bibinfo{author}{\bibfnamefont{W.}~\bibnamefont{Ratcliff~II}},
  \bibinfo{author}{\bibfnamefont{S.~W.} \bibnamefont{Cheong}},
  \bibnamefont{and} \bibinfo{author}{\bibfnamefont{H.~D.} \bibnamefont{Drew}},
  \bibinfo{journal}{Phys. Rev. Lett.} \textbf{\bibinfo{volume}{94}},
  \bibinfo{pages}{137202} (\bibinfo{year}{2005}).

\bibitem[{\citenamefont{Fennie and Rabe}(2006)}]{Fennie2006}
\bibinfo{author}{\bibfnamefont{C.~J.} \bibnamefont{Fennie}} \bibnamefont{and}
  \bibinfo{author}{\bibfnamefont{K.~M.} \bibnamefont{Rabe}},
  \bibinfo{journal}{Phys. Rev. Lett.} \textbf{\bibinfo{volume}{95}},
  \bibinfo{pages}{205505} (\bibinfo{year}{2006}).

\bibitem[{\citenamefont{Rudolf et~al.}(2008)\citenamefont{Rudolf, Kant, Mayr,
  and Loidl}}]{RUDOLF2008}
\bibinfo{author}{\bibfnamefont{T.}~\bibnamefont{Rudolf}},
  \bibinfo{author}{\bibfnamefont{C.}~\bibnamefont{Kant}},
  \bibinfo{author}{\bibfnamefont{F.}~\bibnamefont{Mayr}}, \bibnamefont{and}
  \bibinfo{author}{\bibfnamefont{A.}~\bibnamefont{Loidl}},
  \bibinfo{journal}{Phys. Rev. B} \textbf{\bibinfo{volume}{77}},
  \bibinfo{pages}{024421} (\bibinfo{year}{2008}).

\bibitem[{\citenamefont{Kant et~al.}(2008)\citenamefont{Kant, Rudolf,
  Schrettle, Mayr, Deisenhofer, Lunkenheimer, Eremin, and Loidl}}]{KANT2008}
\bibinfo{author}{\bibfnamefont{C.}~\bibnamefont{Kant}},
  \bibinfo{author}{\bibfnamefont{T.}~\bibnamefont{Rudolf}},
  \bibinfo{author}{\bibfnamefont{F.}~\bibnamefont{Schrettle}},
  \bibinfo{author}{\bibfnamefont{F.}~\bibnamefont{Mayr}},
  \bibinfo{author}{\bibfnamefont{J.}~\bibnamefont{Deisenhofer}},
  \bibinfo{author}{\bibfnamefont{P.}~\bibnamefont{Lunkenheimer}},
  \bibinfo{author}{\bibfnamefont{M.~V.} \bibnamefont{Eremin}},
  \bibnamefont{and} \bibinfo{author}{\bibfnamefont{A.}~\bibnamefont{Loidl}},
  \bibinfo{journal}{Phys. Rev. B} \textbf{\bibinfo{volume}{78}},
  \bibinfo{pages}{245103} (\bibinfo{year}{2008}).

\bibitem[{\citenamefont{Katsumate}(2000)}]{KATSUMATA2000}
\bibinfo{author}{\bibfnamefont{K.}~\bibnamefont{Katsumate}},
  \bibinfo{journal}{J. Phys.: Condens. Matt.} \textbf{\bibinfo{volume}{12}},
  \bibinfo{pages}{R589} (\bibinfo{year}{2000}).

\bibitem[{\citenamefont{Stout and Reed}(1954)}]{STOUT1954}
\bibinfo{author}{\bibfnamefont{J.~W.} \bibnamefont{Stout}} \bibnamefont{and}
  \bibinfo{author}{\bibfnamefont{S.~A.} \bibnamefont{Reed}},
  \bibinfo{journal}{J. Am. Chem. Soc.} \textbf{\bibinfo{volume}{76}},
  \bibinfo{pages}{5279} (\bibinfo{year}{1954}).

\bibitem[{\citenamefont{Lockwood and Cottam}(1988)}]{LOCKWOOD1988}
\bibinfo{author}{\bibfnamefont{D.~J.} \bibnamefont{Lockwood}} \bibnamefont{and}
  \bibinfo{author}{\bibfnamefont{M.~G.} \bibnamefont{Cottam}},
  \bibinfo{journal}{J. Appl. Phys.} \textbf{\bibinfo{volume}{64}},
  \bibinfo{pages}{5876} (\bibinfo{year}{1988}).

\bibitem[{\citenamefont{Weaver et~al.}(1974)\citenamefont{Weaver, Ward,
  Kovener, and Alexander}}]{WEAVER1974}
\bibinfo{author}{\bibfnamefont{J.~H.} \bibnamefont{Weaver}},
  \bibinfo{author}{\bibfnamefont{C.~A.} \bibnamefont{Ward}},
  \bibinfo{author}{\bibfnamefont{G.~S.} \bibnamefont{Kovener}},
  \bibnamefont{and}
  \bibinfo{author}{\bibfnamefont{R.}~\bibnamefont{Alexander}},
  \bibinfo{journal}{J. Phys. Chem. Solids} \textbf{\bibinfo{volume}{35}},
  \bibinfo{pages}{1625} (\bibinfo{year}{1974}).

\bibitem[{\citenamefont{Gibbons}(1959)}]{GIBBONS1959}
\bibinfo{author}{\bibfnamefont{D.~F.} \bibnamefont{Gibbons}},
  \bibinfo{journal}{Phys. Rev.} \textbf{\bibinfo{volume}{115}},
  \bibinfo{pages}{1194} (\bibinfo{year}{1959}).

\bibitem[{\citenamefont{Gervais and Piriou}(1974{\natexlab{a}})}]{Gervais1974}
\bibinfo{author}{\bibfnamefont{F.}~\bibnamefont{Gervais}} \bibnamefont{and}
  \bibinfo{author}{\bibfnamefont{B.}~\bibnamefont{Piriou}},
  \bibinfo{journal}{Phys. Rev. B} \textbf{\bibinfo{volume}{10}},
  \bibinfo{pages}{1642} (\bibinfo{year}{1974}{\natexlab{a}}).

\bibitem[{\citenamefont{Samara and Peercy}(1973)}]{SAMARA1973}
\bibinfo{author}{\bibfnamefont{G.~A.} \bibnamefont{Samara}} \bibnamefont{and}
  \bibinfo{author}{\bibfnamefont{P.~S.} \bibnamefont{Peercy}},
  \bibinfo{journal}{Phys. Rev. B} \textbf{\bibinfo{volume}{7}},
  \bibinfo{pages}{1131} (\bibinfo{year}{1973}).

\bibitem[{\citenamefont{Montanari and Harrison}(2004)}]{MONTANARI2004}
\bibinfo{author}{\bibfnamefont{B.}~\bibnamefont{Montanari}} \bibnamefont{and}
  \bibinfo{author}{\bibfnamefont{N.~M.} \bibnamefont{Harrison}},
  \bibinfo{journal}{J. Phys. Condens. Matt.} \textbf{\bibinfo{volume}{16}},
  \bibinfo{pages}{273} (\bibinfo{year}{2004}).

\bibitem[{\citenamefont{Lockwood et~al.}(1983)\citenamefont{Lockwood, Katiyar,
  and So}}]{LOCKWOOD1983}
\bibinfo{author}{\bibfnamefont{D.~J.} \bibnamefont{Lockwood}},
  \bibinfo{author}{\bibfnamefont{R.~S.} \bibnamefont{Katiyar}},
  \bibnamefont{and} \bibinfo{author}{\bibfnamefont{V.~C.~Y.} \bibnamefont{So}},
  \bibinfo{journal}{Phys. Rev. B} \textbf{\bibinfo{volume}{28}},
  \bibinfo{pages}{1983} (\bibinfo{year}{1983}).

\bibitem[{\citenamefont{Giordano and Sauvajol}(1988)}]{Giordano1988}
\bibinfo{author}{\bibfnamefont{J.}~\bibnamefont{Giordano}} \bibnamefont{and}
  \bibinfo{author}{\bibfnamefont{J.~L.} \bibnamefont{Sauvajol}},
  \bibinfo{journal}{Phys. Stat. Sol. (b)} \textbf{\bibinfo{volume}{147}},
  \bibinfo{pages}{537} (\bibinfo{year}{1988}).

\bibitem[{\citenamefont{Lockwood}(2002)}]{LOCKWOOD2002}
\bibinfo{author}{\bibfnamefont{D.~J.} \bibnamefont{Lockwood}},
  \bibinfo{journal}{Low Temp. Phys.} \textbf{\bibinfo{volume}{28}},
  \bibinfo{pages}{505} (\bibinfo{year}{2002}).

\bibitem[{\citenamefont{Homes et~al.}(1993)\citenamefont{Homes, Reedyk,
  Crandles, and Timusk}}]{HOMES1993b}
\bibinfo{author}{\bibfnamefont{C.~C.} \bibnamefont{Homes}},
  \bibinfo{author}{\bibfnamefont{M.}~\bibnamefont{Reedyk}},
  \bibinfo{author}{\bibfnamefont{D.~A.} \bibnamefont{Crandles}},
  \bibnamefont{and} \bibinfo{author}{\bibfnamefont{T.}~\bibnamefont{Timusk}},
  \bibinfo{journal}{Appl. Opt.} \textbf{\bibinfo{volume}{32}},
  \bibinfo{pages}{2976} (\bibinfo{year}{1993}).

\bibitem[{\citenamefont{Dovesi et~al.}(2006)\citenamefont{Dovesi, Saunders,
  Roetti, Orlando, Zicovich-Wilson, Pascale, Civalleri, Doll, Harrison, Bush
  et~al.}}]{CRYSTAL}
\bibinfo{author}{\bibfnamefont{R.}~\bibnamefont{Dovesi}},
  \bibinfo{author}{\bibfnamefont{V.~R.} \bibnamefont{Saunders}},
  \bibinfo{author}{\bibfnamefont{C.}~\bibnamefont{Roetti}},
  \bibinfo{author}{\bibfnamefont{R.}~\bibnamefont{Orlando}},
  \bibinfo{author}{\bibfnamefont{C.~M.} \bibnamefont{Zicovich-Wilson}},
  \bibinfo{author}{\bibfnamefont{F.}~\bibnamefont{Pascale}},
  \bibinfo{author}{\bibfnamefont{B.}~\bibnamefont{Civalleri}},
  \bibinfo{author}{\bibfnamefont{K.}~\bibnamefont{Doll}},
  \bibinfo{author}{\bibfnamefont{N.~M.} \bibnamefont{Harrison}},
  \bibinfo{author}{\bibfnamefont{I.~J.} \bibnamefont{Bush}},
  \bibnamefont{et~al.}, \emph{\bibinfo{title}{CRYSTAL06 User's manual}}
  (\bibinfo{publisher}{University of Torino}, \bibinfo{year}{2006}).

\bibitem[{\citenamefont{Becke}(1993)}]{B3LYP}
\bibinfo{author}{\bibfnamefont{A.~D.} \bibnamefont{Becke}},
  \bibinfo{journal}{J. Chem. Phys.} \textbf{\bibinfo{volume}{98}},
  \bibinfo{pages}{5648} (\bibinfo{year}{1993}).

\bibitem[{\citenamefont{Bilc et~al.}(2008)\citenamefont{Bilc, Orlando, Shaltaf,
  Rignanese, \'I\~niguez, and Ghosez}}]{B1PW}
\bibinfo{author}{\bibfnamefont{D.~I.} \bibnamefont{Bilc}},
  \bibinfo{author}{\bibfnamefont{R.}~\bibnamefont{Orlando}},
  \bibinfo{author}{\bibfnamefont{R.}~\bibnamefont{Shaltaf}},
  \bibinfo{author}{\bibfnamefont{G.~M.} \bibnamefont{Rignanese}},
  \bibinfo{author}{\bibfnamefont{J.}~\bibnamefont{\'I\~niguez}},
  \bibnamefont{and} \bibinfo{author}{\bibfnamefont{P.}~\bibnamefont{Ghosez}},
  \bibinfo{journal}{Phys. Rev. B} \textbf{\bibinfo{volume}{77}},
  \bibinfo{pages}{165107} (\bibinfo{year}{2008}).

\bibitem[{\citenamefont{Martin and Sundermann}(2001)}]{BaseMn}
\bibinfo{author}{\bibfnamefont{J.}~\bibnamefont{Martin}} \bibnamefont{and}
  \bibinfo{author}{\bibfnamefont{A.}~\bibnamefont{Sundermann}},
  \bibinfo{journal}{J. Chem. Phys.} \textbf{\bibinfo{volume}{114}},
  \bibinfo{pages}{3408} (\bibinfo{year}{2001}), \bibinfo{note}{diffuse orbitals
  with exponents samller than 0.15 were omitted as usual in infinite systems.}

\bibitem[{\citenamefont{Pseudo~:~Dolg et~al.}(1987)\citenamefont{Pseudo~:~Dolg,
  Wedig, Stoll, and Preuss}}]{PseudoMn}
\bibinfo{author}{\bibfnamefont{M.}~\bibnamefont{Pseudo~:~Dolg}},
  \bibinfo{author}{\bibfnamefont{U.}~\bibnamefont{Wedig}},
  \bibinfo{author}{\bibfnamefont{H.}~\bibnamefont{Stoll}}, \bibnamefont{and}
  \bibinfo{author}{\bibfnamefont{H.}~\bibnamefont{Preuss}},
  \bibinfo{journal}{J. Chem. Phys.} \textbf{\bibinfo{volume}{86}},
  \bibinfo{pages}{866} (\bibinfo{year}{1987}).

\bibitem[{\citenamefont{Prencipe et~al.}(1995)\citenamefont{Prencipe, Zupan,
  Dovesi, Apr\`a , and Saunders}}]{BaseF}
\bibinfo{author}{\bibfnamefont{M.}~\bibnamefont{Prencipe}},
  \bibinfo{author}{\bibfnamefont{A.}~\bibnamefont{Zupan}},
  \bibinfo{author}{\bibfnamefont{R.}~\bibnamefont{Dovesi}},
  \bibinfo{author}{\bibfnamefont{R.}~\bibnamefont{Apr\`a }}, \bibnamefont{and}
  \bibinfo{author}{\bibfnamefont{V.~R.} \bibnamefont{Saunders}},
  \bibinfo{journal}{Phys. Rav. B} \textbf{\bibinfo{volume}{51}},
  \bibinfo{pages}{3391} (\bibinfo{year}{1995}).

\bibitem[{\citenamefont{Sun et~al.}(2008)\citenamefont{Sun, Allen, Stahnke,
  Jacobsen, and Homes}}]{Sun2008}
\bibinfo{author}{\bibfnamefont{T.}~\bibnamefont{Sun}},
  \bibinfo{author}{\bibfnamefont{P.~B.} \bibnamefont{Allen}},
  \bibinfo{author}{\bibfnamefont{D.~G.} \bibnamefont{Stahnke}},
  \bibinfo{author}{\bibfnamefont{S.~D.} \bibnamefont{Jacobsen}},
  \bibnamefont{and} \bibinfo{author}{\bibfnamefont{C.~C.} \bibnamefont{Homes}},
  \bibinfo{journal}{Phys. Rev. B} \textbf{\bibinfo{volume}{77}},
  \bibinfo{pages}{134303} (\bibinfo{year}{2008}).

\bibitem[{\citenamefont{Benoit and Giordano}(1988)}]{Benoit1988}
\bibinfo{author}{\bibfnamefont{C.}~\bibnamefont{Benoit}} \bibnamefont{and}
  \bibinfo{author}{\bibfnamefont{J.}~\bibnamefont{Giordano}},
  \bibinfo{journal}{J. Phys. C: Solid State Phys.}
  \textbf{\bibinfo{volume}{21}}, \bibinfo{pages}{5209} (\bibinfo{year}{1988}).

\bibitem[{\citenamefont{Gervais and Piriou}(1974{\natexlab{b}})}]{Gervais1974b}
\bibinfo{author}{\bibfnamefont{F.}~\bibnamefont{Gervais}} \bibnamefont{and}
  \bibinfo{author}{\bibfnamefont{B.}~\bibnamefont{Piriou}},
  \bibinfo{journal}{J. Phys. C} \textbf{\bibinfo{volume}{7}},
  \bibinfo{pages}{2374} (\bibinfo{year}{1974}{\natexlab{b}}).

\bibitem[{\citenamefont{Kuzmenko}(2005)}]{KUZMENKO2005}
\bibinfo{author}{\bibfnamefont{A.~B.} \bibnamefont{Kuzmenko}},
  \bibinfo{journal}{Rev. Sci. Instrum.} \textbf{\bibinfo{volume}{76}}
  (\bibinfo{year}{2005}).

\bibitem[{\citenamefont{Baltensperger and Helman}(1968)}]{Baltensperger1968}
\bibinfo{author}{\bibfnamefont{W.}~\bibnamefont{Baltensperger}}
  \bibnamefont{and} \bibinfo{author}{\bibfnamefont{J.~S.}
  \bibnamefont{Helman}}, \bibinfo{journal}{Helv. Phys. Acta}
  \textbf{\bibinfo{volume}{41}}, \bibinfo{pages}{668} (\bibinfo{year}{1968}).

\bibitem[{\citenamefont{Perakis et~al.}(1999)\citenamefont{Perakis,
  Sarantopoulou, Raptis, and Raptis}}]{PERAKIS1999}
\bibinfo{author}{\bibfnamefont{A.}~\bibnamefont{Perakis}},
  \bibinfo{author}{\bibfnamefont{E.}~\bibnamefont{Sarantopoulou}},
  \bibinfo{author}{\bibfnamefont{Y.~S.} \bibnamefont{Raptis}},
  \bibnamefont{and} \bibinfo{author}{\bibfnamefont{C.}~\bibnamefont{Raptis}},
  \bibinfo{journal}{Phys. Rev. B} \textbf{\bibinfo{volume}{59}},
  \bibinfo{pages}{775} (\bibinfo{year}{1999}).

\bibitem[{\citenamefont{Perakis et~al.}(2005)\citenamefont{Perakis, Lampakis,
  Boulmetis, and Raptis}}]{PERAKIS2005}
\bibinfo{author}{\bibfnamefont{A.}~\bibnamefont{Perakis}},
  \bibinfo{author}{\bibfnamefont{D.}~\bibnamefont{Lampakis}},
  \bibinfo{author}{\bibfnamefont{Y.~C.} \bibnamefont{Boulmetis}},
  \bibnamefont{and} \bibinfo{author}{\bibfnamefont{C.}~\bibnamefont{Raptis}},
  \bibinfo{journal}{Phys. Rev. B} \textbf{\bibinfo{volume}{72}},
  \bibinfo{pages}{144108} (\bibinfo{year}{2005}).

\bibitem[{\citenamefont{Yamaguchi et~al.}(1992)\citenamefont{Yamaguchi, Yagi,
  and Hamaya}}]{YAMAGUCHI1992}
\bibinfo{author}{\bibfnamefont{M.}~\bibnamefont{Yamaguchi}},
  \bibinfo{author}{\bibfnamefont{T.}~\bibnamefont{Yagi}}, \bibnamefont{and}
  \bibinfo{author}{\bibfnamefont{N.}~\bibnamefont{Hamaya}},
  \bibinfo{journal}{J. Phys. Soc. Japan} \textbf{\bibinfo{volume}{61}},
  \bibinfo{pages}{3883} (\bibinfo{year}{1992}).

\bibitem[{\citenamefont{Melcher}(1970)}]{Melcher1970}
\bibinfo{author}{\bibfnamefont{R.~L.} \bibnamefont{Melcher}},
  \bibinfo{journal}{Phys. Rev. B} \textbf{\bibinfo{volume}{2}},
  \bibinfo{pages}{733} (\bibinfo{year}{1970}).

\bibitem[{\citenamefont{Rimai}(1977)}]{Rimai1977}
\bibinfo{author}{\bibfnamefont{D.~S.} \bibnamefont{Rimai}},
  \bibinfo{journal}{Phys. Rev. B} \textbf{\bibinfo{volume}{16}},
  \bibinfo{pages}{4069} (\bibinfo{year}{1977}).

\bibitem[{\citenamefont{Giordano and Sauvajol}(1987)}]{Giordano1987}
\bibinfo{author}{\bibfnamefont{J.}~\bibnamefont{Giordano}} \bibnamefont{and}
  \bibinfo{author}{\bibfnamefont{J.~L.} \bibnamefont{Sauvajol}},
  \bibinfo{journal}{J. Phys. C: Solid State Phys.}
  \textbf{\bibinfo{volume}{20}}, \bibinfo{pages}{1547} (\bibinfo{year}{1987}).

\bibitem[{\citenamefont{Seehra et~al.}(1986)\citenamefont{Seehra, Helmick, and
  Srinivasan}}]{SEEHRA1986}
\bibinfo{author}{\bibfnamefont{M.~S.} \bibnamefont{Seehra}},
  \bibinfo{author}{\bibfnamefont{R.~E.} \bibnamefont{Helmick}},
  \bibnamefont{and}
  \bibinfo{author}{\bibfnamefont{G.}~\bibnamefont{Srinivasan}},
  \bibinfo{journal}{J. Phys. C: Solid State Phys.}
  \textbf{\bibinfo{volume}{19}}, \bibinfo{pages}{1627} (\bibinfo{year}{1986}).

\bibitem[{\citenamefont{Ure}(1957)}]{Ure1957}
\bibinfo{author}{\bibfnamefont{R.~W.} \bibnamefont{Ure}}, \bibinfo{journal}{J.
  Chem. Phys.} \textbf{\bibinfo{volume}{26}}, \bibinfo{pages}{1363}
  (\bibinfo{year}{1957}).

\bibitem[{\citenamefont{Samara and Peercy}(1979)}]{SAMARA1979}
\bibinfo{author}{\bibfnamefont{G.~A.} \bibnamefont{Samara}} \bibnamefont{and}
  \bibinfo{author}{\bibfnamefont{P.~S.} \bibnamefont{Peercy}},
  \bibinfo{journal}{J. Phys. Chem. Solids} \textbf{\bibinfo{volume}{40}},
  \bibinfo{pages}{509} (\bibinfo{year}{1979}).

\bibitem[{\citenamefont{Momma and Izumi}(2008)}]{VESTA}
\bibinfo{author}{\bibfnamefont{K.}~\bibnamefont{Momma}} \bibnamefont{and}
  \bibinfo{author}{\bibfnamefont{F.}~\bibnamefont{Izumi}}, \bibinfo{journal}{J.
  Appl. Crystallogr.} \textbf{\bibinfo{volume}{41}}, \bibinfo{pages}{653}
  (\bibinfo{year}{2008}).

\end{thebibliography}

\end{document}